\documentclass[aps,twocolumn,superscriptaddress]{revtex4}

\usepackage{amsmath,amssymb,bm}
\usepackage{graphicx}
\usepackage{amsfonts}
\usepackage{color}
\usepackage{color}

\graphicspath{{./figs/}}

\usepackage[abs]{overpic}

\newcommand{\nn}{\nonumber}                                           
\newcommand{\va}[1]{\langle{#1}\rangle}                               

\newcommand{\al}{\alpha}
\newcommand{\be}{\beta}
\newcommand{\ga}{\gamma}
\newcommand{\de}{\delta}

\newcommand{\la}{\lambda}
\newcommand{\ro}{\rho}
\newcommand{\si}{\sigma}

\newcommand{\Tr}{\mathop{\rm Tr}\nolimits}

\newcommand{\mycomment}[1]{\iffalse #1 \fi}

\begin{document}

\date{\today}

\title{Nonlocal gluon condensates in QCD sum rules}

\author{Alexandr V. Pimikov}
\email{pimikov@gmail.com}
\affiliation{Bogoliubov Laboratory of Theoretical Physics, Joint
    Institute for Nuclear Research,\\ Dubna, Moscow Region, 141980
    Russia}

\begin{abstract}

	Nonlocal gluon condensates are vacuum expectations of the product of gluon field strength tensors.
Short-distance expansions of two-, three-, and four-gluon condensates are presented up to dimension-8 local operators.
We propose a method for calculating the Wilson coefficients based on the presented expansions and the Feynman diagram
technique in the background field approach.
The method is demonstrated using the glueball current correlators as examples.
Methodological aspects of the background field approach are discussed in relation to glueball studies within QCD sum rules.
We confirm the results for Operator Product Expansion (OPE) of the two-gluon $0^{\pm +}$ glueball current correlators
and calculate additional contributions coming from dimension-6 four-quark condensate and dimension-8 mixed quark-gluon condensates.
The OPEs used in QCD sum rules for three-gluon $0^{\pm +}$ glueballs are revisited up to dimension-6 order.

\end{abstract}

\maketitle

\section{Introduction}

A practical method to evaluate quantitatively the physical characteristics of hadrons from the QCD is provided by an approach called the QCD Sum Rules (SR)~\cite{Shifman:1978bw,Shifman:1978bx,Shifman:1978by}.
This approach gives a the direct correspondence between the hadron parameters and Wilson's operator-product expansion (OPE)~\cite{Wilson:1969zs} of correlation functions. 
The correspondence arises from the dispersion relations.
Various aspects and reviews of QCD SRs applications can 
be found in the following Refs.~\cite{Novikov:1980uj,Narison:1980ti,Novikov:1981xi,Chernyak:1983ej,Novikov:1983gd,Shuryak:1984nq,Reinders1985,Shifman:1993wf,Cohen:1994wm,Colangelo:2000dp,Khodjamirian:2002pka,Narison:2002pw,Hofmann:2003qf,Narison:2005hb,Nielsen:2009uh,Narison:2010wb,Shifman:2010zzb,Narison:2014wqa,Chen:2016qju,Ioffe:2010zz,Dominguez:2018zzi,Khodjamirian2020book, Hala:2021sij},
for recent development of QCD SR in medium, see~\cite{Gubler:2018ctz,Kim:2017nyg,Buchheim:2015yyc}.
Sum rules are useful beyond the Standard model in application for bounds on New Physics effects~\cite{Shifman:1979if,Osamura:2022rak,Ema:2022pmo}.

The OPE terms are given as a product of the coefficient function and the vacuum condensate that reflect the separation of short and large distance effects, respectively. 
The approaches to calculate the coefficient functions include 
the method of projectors~\cite{Gorishnii:1983su,Gorishnii:1986gn}, 
Schwinger's operator method~\cite{Schwinger:1951nm} and its branch based on the Fock-Schwinger gauge,
and the background field approach, see references  		
in~\cite{Novikov:1983gd}.

The technical details of calculating the coefficient function are usually omitted
due to complexity. 
A consistent procedure for such calculations was elaborated in~\cite{Novikov:1983gd} and further developed in~\cite{Grozin:1994hd}.
The presented study can be considered as a continuation of~\cite{Grozin:1994hd}.
We propose a method for calculating the coefficient function based on the nonlocal condensates (NLCs)~\cite{Gromes:1982su,Mikhailov:1988nz,Bakulev:2001pa}
and Feynman diagram technique in the framework of the background field approach.
Nonlocal gluon condensates are vacuum expectations of the product of gluon field
strength tensors.
For example, the two-gluon nonlocal condensate has dimension four in energy units (dimension-4) and can be defined as ``physical vacuum'' expectation of the normal product
$\va{0|:G_{\mu\nu}(x)[x;y]G_{\al\be}(y):|0}$, where~$G_{\mu\nu}$ is the gluon field strength tensor and $[x;y]$ is a Wilson line, see details in Sec.~\ref{sec:2Gluon.cond}. 

The methodological aspects of QCD SR are often considered for the case of quark current correlators, see~\cite{Albuquerque:2013ija}, while the glueball currents correlators have particularities left unattended.
Therefore, applications of the proposed approach are given by the examples of 
the glueball currents correlators used in QCD SRs studies~\cite{Novikov:1979ux,Novikov:1979va,Latorre:1987wt,Hao:2005hu}.

Glueballs are an exotic state that contains only gluons and no valence quarks.
This type of state has candidates among observations~\cite{Klempt:2007cp,Ochs:2013gi,Jia:2016cgl,Klempt:2022qjf,Csorgo:2019ewn} and is included in the running and projected large-scale experiments:
Belle (Japan),  BESIII (Beijing, China),
LHC (CERN), GlueX (JLAB,USA),
NICA (Dubna, Russia), HIAF (China), and FAIR (GSI, Germany). Electorn-Ion-Colliders have potential for glueball state observations~\cite{Chekanov:2022sax}.
The reviews of glueball physics can be found in~\cite{Mathieu:2008me,Ochs:2013gi,Chen:2022asf}. 
There are many applications of QCD SR to glueball states%
~\cite{Novikov:1981xi,Novikov:1979ux,Novikov:1979va,Novikov:1979uy,Novikov:1980dj,Shifman1981,Krasnikov:1982ea,Narison:1984hu,Novikov:1984rf,Bordes:1989kc,Narison:1988ts,Bagan:1990sy,Wakely:1991eu,Narison:1996fm,Liu:1998xx,Huang1999,Forkel:2000fd,Harnett:2000fy,Zhang:2003mr,Forkel:2003mk,Narison:2005wc,Xian:2014jpa,Wang:2015kra,Pimikov:2017bkk,Pimikov:2017xap,Pimikov:2016pag,Chen:2017ror,Latorre:1987wt,Hao:2005hu,Qiao:2014vva,Qiao:2015iea,Chen:2021cjr,Chen:2021bck}.
Here we shortly discuss some of them.
The first QCD SR study~\cite{Novikov:1979ux} of glueballs considered a pseudo-scalar $0^{-+}$ state
with an obtained mass of $\sim 1$~GeV, where two-gluon current was used.
Later QCD SR was applied~\cite{Novikov:1979va} to a scalar $0^{++}$ glueball state, where the glueball mass was estimated to be $\sim 0.7$~GeV.
The OPE used in SRs for the scalar and pseudoscalar glueballs was extended~\cite{Harnett:2000fy,Forkel:2003mk,Zoller:2013ixa,Zhang:2003mr} by 
including the direct instanton contribution,  
the two loop radiative corrections~\cite{Kataev:1981gr,Kataev:1981aw} to the perturbative term,
and the one loop radiative correction~\cite{Bagan:1989vm} to dimension-4 term (dimension of the considered condensate is four in energy units).
Three-gluon glueballs were first studied in~\cite{Latorre:1987wt} for a $0^{++}$-state. Then, QCD SRs for three-gluon glueballs was extended~\cite{Hao:2005hu} to the $0^{-+}$ scalar, vector and tensor states~\cite{Liu:1998xx}.

Here we study the technical aspects of the calculation of OPE's coefficients up to dimension-8 order
and suggest a new way to organize and perform calculations using 
NLCs~\cite{Gromes:1982su,Mikhailov:1988nz,Bakulev:2001pa}.
Applying the algorithms formulated in~\cite{Grozin:1994hd} for calculating  higher power corrections, we extend the results of~\cite{Grozin:1994hd} to a full set of gluon NLCs
needed for calculations of OPE up to dimension-8 terms without considering the radiative corrections to the coefficient functions, see Sec.~\ref{sec:nlc}. 
The three-loop coefficient functions of two-gluon NLC OPE was obtained in~\cite{Braun:2021cqe} for the leading dimension-4 term.
We develop and obtain expansions of NLCs for two and three-gluon condensate up to dimension-8,
where terms of expansions are given employing local condensates. 
Vacuum expectation of normal product of local operators are usually called condensate
but here we use phrase ``local condensate'' to distinguish between local and nonlocal condensate.
The result for four-gluon condensate expansion includes only the leading dimension-8 term.
Coordinate dependence of higher dimension terms
is also discussed using permutation symmetries of gluon fields strength tensor in condensate.
Our result is in agreement with the two-gluon condensate expansions that were first obtained in~\cite{Mikhailov:1992ug} in dimension-6 order and then in~\cite{Grozin:1994hd} up to dimension-8 term.

The number of terms that need to be evaluated is growing with the mass dimension
(dimension in energy units) of  OPE order. 
We propose~\cite{Pimikov:2016pag,Pimikov:2017xap,Pimikov:2017bkk} to use the so-called nonlocal condensates, which help to systematize the contributions and simplify calculations. 
We consider NLCs in the form of their truncated expansions given in terms of local condensates.
As a result, each NLC-based OPE contribution is represented in this work through a finite set of the OPE terms given by the local condensates of various dimensions starting from the mass dimension of the original NLC. 
In other words, we use NLCs only as intermediate while the final results for OPEs are given through local condensates.
Originally, NLCs were used differently in QCD SR -- for resumming
an infinite series of local condensates by modeling of the long-range dependence~\cite{Mikhailov:1991pt} of NLCs.
One of the models for the two-gluon NLC could be found in~\cite{Dorokhov:1999ig}.
The NLCs were successfully applied in studies of the hadron structures (distribution amplitudes, form factors)~\cite{Mikhailov:1986be,Mikhailov:1991pt,Grozin:1992td,Mikhailov:1992ug,Bakulev:2001pa,Mikhailov:2010ud,Pimikov:2008ay,Pimikov:2009mq,Bakulev:2009hi,Bakulev:2009ib,Pimikov:2013usa,Pimikov:ppnl}. 

The idea of using a truncated series of NLCs for OPE was first applied in~\cite{Pimikov:2016pag,Pimikov:2017xap,Pimikov:2017bkk} for studying C-odd $0^{\pm-}$ three-gluon glueball states in QCD SR. 
Although,  we focus here on the  gluon condensates 
and glueball state, the considered ideas 
can be applied to other states within QCD SR and beyond.
Recent applications of gluon condensates beyond QCD SR include studies of heavy quarkonium within potential nonrelativistic QCD~\cite{Brambilla:2020ojz,Brambilla:2020xod}, the rapidity anomalous dimension or Collins-Soper kernel~\cite{Vladimirov:2020umg}, 
see references in~\cite{Braun:2021cqe}.

The suggested method to calculate OPE using a truncated series of NLCs is demonstrated for correlators that represent the two-gluon and three-gluon $0^{\pm+}$ glueball states in QCD SRs.
The  demonstrations lead to the following results: (i)
for the two-gluon current cases we obtained additional dimension-8 terms 
which were not considered in~\cite{Novikov:1979ux,Novikov:1979va};
(ii) for the three-gluon current cases~\cite{Latorre:1987wt,Hao:2005hu} the OPE for correllators are revisited in dimension-6 order and compared with~\cite{Latorre:1987wt,Hao:2005hu}.

The paper is organized as follows. In the next section, the explicit expressions for tensor and Taylor expansions of gluon NLCs are presented in terms of local condensates. 
Then, in Sec. \ref{sec:usage}, we apply these expansions to correlators
of glueball currents.
In Sec. \ref{sec:usage}, we also discuss aspects of the background field approach, especially those that are relevant to glueball studies within QCD SR.
In Sec.~\ref{sec:conclusions}, we summarize our observations and results.
In the Appendices, we provide the Feynman rules and a detailed example of their application.

\section{Gluon condensates}
\label{sec:nlc}

In this section, we present the expansion of two-, \linebreak three-, and four-gluon NLCs in terms of local condensates. The expansions are obtained by the method formulated in~\cite{Grozin:1994hd}. 
The coefficient functions and operators in OPE are gauge invariant; therefore,
the choice of gauge is a matter of convenience.
The expansion of NLCs is performed in the  FS (Fock--Schwinger) gauge~\footnote{The FS gauge is also known as a fixed-point gauge, radial gauge, and coordinate gauge.}:
\begin{equation}
	(x_\mu-z_\mu) A^a_\mu(x)=0\,,
	\label{eq:FPG}
\end{equation}
with the gauge-fixing point $z_\mu=0$.
In the FS gauge the Taylor expansion for the gluon field strength tensor $G^a_\mu(x)$ can be
written in gauge--covariant form~\cite{Novikov:1983gd}:
\begin{eqnarray}\label{eq:Gexpansion}
	G^a_{\mu\nu}(x)&=&G^a_{\mu\nu}(0)
	+\frac1{1!}x_\al D_\al G^a_{\mu\nu}(0)
	\\\nn&&+\frac1{2!}x_\be x_\al  D_\be D_\al G^a_{\mu\nu}(0)
	+O(x^3)\,.
	\label{eq:FPGexp}
\end{eqnarray}
Here we use the covariant derivative $D_\mu G^a_{\rho\sigma}=(\partial_\mu\delta^{ab}+gf^{acb}A^c_\mu)G^b_{\rho\sigma}$ 
in the adjoint representation.
The advantage of FS gauge is that, due to Eq.~(\ref{eq:FPG}), the gluon field can be expressed via the field strength tensor:
\begin{eqnarray}\label{eq:FS-integral} 
	A^a_\nu(x)=z_\mu\!\int_0^1\!\!dt~t~G^a_{\mu\nu}(tx)\,.
\end{eqnarray}
Taylor's expansion of the gluon field can be obtained from Eq.~(\ref{eq:FS-integral})
and Eq.~(\ref{eq:Gexpansion}).
Using the above equations  (\ref{eq:Gexpansion}) and  (\ref{eq:FS-integral}), one can expand 
gluon NLCs in terms of local condensates. 

The reader may be familiar with the following notation for local condensates
\footnote{Across the work the condensate $\va{O}$ of any operator $O$ is defined as vacuum expectation of the  normal product $\va{O}\equiv \va{0|:O:|0}$.}:
one dimension-4 condensate 
$\va{\alpha_SG^2}\equiv \va{\alpha_S G^a_{\mu\nu}G^a_{\mu\nu}}$,
two dimension-6 condensates
$\va{gG^3}\equiv \va{gf^{abc}G^a_{\mu\nu}G^b_{\nu\rho}G^c_{\rho\mu}}$,
$\va{J^2}\equiv \va{J^a_\mu J^a_\mu}$, 
and two dimension-8 condensates
$\va{(f^{abc}G^b_{\mu\nu}G^c_{\alpha\beta})^2}$,
$\va{(f^{abc}G^b_{\mu\nu}G^c_{\nu\rho})^2}$, where $g$ is the coupling constant of strong interaction with $\al_S=g^2/(4\pi)$ and the quark current $J_\mu^a=\sum \bar q\ga_\mu t^a q$.
We use a different notation for local gluon condensates suggested in~\cite{Broadhurst:1985js,Grozin:1994hd} that includes more condensates at dimension-8 order. Gluon condensate of dimension-4
and condensates of dimension-6 are defined as follows:
\begin{eqnarray} 	\label{eq:G6}
	&&G^4=\va{\Tr G_{\mu\nu}G_{\mu\nu}}\,,
	G^6_1=i\va{\Tr G_{\la\mu}G_{\mu\nu}G_{\nu\la}},
\\\nn &&
	G^6_2=\va{\Tr J_\mu J_\mu}\,,
\end{eqnarray}
where we use compact matrix notation 
$G_{\mu\nu}\equiv  g G_{\mu\nu}^a(0) t^a$ for the gluon field strength tensor and  
$J_\mu\equiv g J_\mu^a t^a$ for the quark current.
A set of independent dimension-8 gluon condensates was found in~\cite{Nikolaev:1982ra}.
Here we use the notation introduced in~\cite{Broadhurst:1985js,Grozin:1986xh}:
\begin{eqnarray}
	&&G^8_1=\va{\Tr G_{\mu\nu}G_{\mu\nu}G_{\al\be}G_{\al\be}},
	\nonumber\\
	&&
	G^8_2=\va{\Tr G_{\mu\nu}G_{\al\be}G_{\mu\nu}G_{\al\be}},
	\nonumber\\
	&&G^8_3=\va{\Tr G_{\mu\al}G_{\al\nu}G_{\nu\be}G_{\be\mu}}, 
		\nonumber\\
	&&G^8_4=\va{\Tr G_{\mu\al}G_{\al\nu}G_{\mu\be}G_{\be\nu}},
	\label{eq:G8}\\
	&&G^8_5=i\va{\Tr J_\mu G_{\mu\nu}J_\nu}, \quad
	\nonumber\\
	&&
	G^8_6=i\va{\Tr J_\la [D_\la G_{\mu\nu},G_{\mu\nu}]},
	\nonumber\\\nonumber
	&&G^8_7=\va{\Tr J_\mu D^2J_\mu}\,,
\end{eqnarray}
where the condensate $G^8_7$ is defined in a different way as suggested in~\cite{Grozin:1994hd}. The notation $G^i_j$ specifies the local condensate
of dimension-$i$.
The trace is taken in the fundamental representation 
with the covariant derivative $D_\mu=\partial_\mu-iA_\mu$ and $A_\mu=gA^a_\mu t^a$.
We observe that most of the dimension-8 terms of gluon NLC expansions are expressed only by
four linear combinations of seven condensates $G^8_i$ ($i=1,\cdots, 7$):
\begin{eqnarray}\label{eq:G8add}
	&&G^8_{12}=G^8_1-G^8_2\,,~~G^8_{34}=G^8_3-G^8_4\,,
	\\\nn &&
	G^8_{56}=4G^8_5-G^8_6\,,~~G^8_{67}=6G^8_7-G^8_6\,.
\end{eqnarray}
There are the following relations with common notations:
\begin{eqnarray}\nn
	G^4_1&=&2\pi \va{\alpha_SG^2}\,,~ 
	\\\nn
	G^6_1&=&-g^2\va{gG^3}/4\,,~
	G^6_2=g^2\va{J^2}/2\,,~\\\nn
	G^8_{12}&=&g^4\va{(f^{abc}G^b_{\mu\nu}G^c_{\alpha\beta})^2}/4\,,~
	\\\nn
	G^8_{34}&=&g^4\va{(f^{abc}G^b_{\mu\nu}G^c_{\nu\rho})^2}/4\,.
\end{eqnarray}
The combinations $G^8_{56}$,$G^8_{67}$ have no commonly used notation.

As we will see in the next subsection, the expansion of gluon fields provides
for NLCs up to dimension-8
not only the gluon condensates $G^4$, $G^6_1$,
$G^8_{12}$, $G^8_{34}$
but also the four quark condensate $G^6_2$ and mixed quark-gluon condensate  $G^8_5$, $G^8_6$, $G^8_7$.
The quark and mixed quark-gluon condensates have not been considered in glueball studies, including~\cite{Novikov:1979ux,Novikov:1979va}. 
In instanton models, these condensates are equal to zero due to the self-duality of the vacuum gluon field strength tensor. 
One may expect that the condensates $G^6_2$, $G^8_5$, $G^8_6$, $G^8_7$  could give a numerically minor correction compared to the pure gluon condensates $G^6_1$,
$G^8_{12}$, $G^8_{34}$. 
In the general case, these condensates should be included in the OPE  as they could be enhanced by the coefficient function.

In the next subsection, we present expansions of nonlocal
gluon condensates in terms of local condensates.
Expanding NLCs at dimension-8 order requires tedious calculation 
described in ~\cite{Grozin:1994hd}.
The obtained expansions are one of the important results of this work.

\begin{table}[h]
	\caption{\label{tab:condGdim6tensorProperties}
		The properties of the tensors $\Gamma_{n}(x,y)$ for the expansion of the two-gluon NLC. 
		The first row specifies the tensor. 
		The second row gives the leading dimension
		of the condensate that contributes to the tensor. The third row tells us 
		which of the tensors contribute in the case when one of the gluon coordinates coincides with the gauge fixing point, e.g., $x=0$ or $y=0$.
		The fourth row demonstrates which of the tensors is nonzero in the collinear case when the coordinates  $x$ and $y$  lying on one line with the  gauge fixing point $x=t y$.
	}
	\begin{center}
		{			
				\begin{tabular}{ccccccccc}\hline\hline
					n   & 0 & 1 & ~~2~~ & 3 & ~~4~~ & ~~5~~ & ~~6~~ & ~~7~~ \\\hline
					dimension & 4 & 6 & 6 & 8 & 8 & 8 & 8 & 8  \\ 
					{\normalsize $x=0$}  & ~~$\neq 0$~~ & ~~$\neq 0$~~ & 0 & 0 & 0 & 0 & 0 & 0  \\
					{\normalsize $x=ty$} & $\neq 0$ & $\neq 0$ & 0 & ~~$\neq 0$~~ & 0 & 0 & 0 & 0  \\\hline\hline
				\end{tabular}
		}
	\end{center}
\end{table}

\subsection{Two-gluon NLC}
\label{sec:2Gluon.cond}

The two-gluon NLC expansion was first presented in dimension-6 order~\cite{Mikhailov:1992ug}, where only the collinear part of the condensate was considered. 
Later, expansion in the non-collinear case was obtained in~\cite{Grozin:1994hd} up to dimension-8 term.
The result obtained here  is in agreement with the expansion given in~\cite{Grozin:1994hd}.
We suggest using the following form for two-gluon NLCs:
\begin{widetext}
\begin{eqnarray}\label{eq:GGnlc}
	\va{\Tr G_{\mu_1\nu_1}(x)[x;y] G_{\mu_2\nu_2}(y)[y;x]}=
	\frac{1}
	{d(d-1)}
	\mathbb{A}({\mu_1,\nu_1})\mathbb{A}({\mu_2,\nu_2})
	\sum\limits_{k=0}^{8}
	\Gamma_{k}(x,y) M_{k}(x,y)\,,
\end{eqnarray}
\end{widetext}
where $d$ is the space dimension, and Wilson's line $[x;y]$ insures gauge invariance and is defined by the P-ordered exponent:
\begin{eqnarray}\nn
	[x;y]=
	\mathbb{P}\text{exp} \Big\{ ig\int_{P(x,y)} \!\!\! d\omega_\mu A_\mu(\omega)\Big\}\,.
\end{eqnarray}

Here and below we use the operator $\mathbb{A}({\mu,\nu})$ for antysymmetrization: $\mathbb{A}({\mu,\nu}) t_{\mu\nu} \equiv t_{\mu\nu}-t_{\nu\mu}$.
The suggested form of expansion, Eq.~(\ref{eq:GGnlc}), is more appropriate for  modeling scalar functions than the one given in~\cite{Grozin:1994hd} and explicitly  displays the symmetries of the condensate with 
respect to the Lorentz indices of strength tensors and with respect to the transformation $\mu_1\leftrightarrow \mu_2$, $\nu_1\leftrightarrow \nu_2$, and $x\leftrightarrow y$ related to the symmetry of gluons field strength tensor permutations in the condensate. 
The new basis leads to simpler expansions of the scalar functions $M_k$ that are easier for modeling long distance behavior of NLC. 
The path $P(x,y)$ is chosen to be a broken line with apexes at the points $x,0,y$; therefore, the links can be omitted $[x;y]=[x;0][0;y]=1$ due to the gauge condition Eq. (\ref{eq:FPG}).
The same path will be applied to three-gluon and four-gluon condensates.
The operators of asymmetrizaton 
$\mathbb{A}(\mu_1,\nu_1)$ allow one to introduce a brief notation for master tensors:
\begin{eqnarray}\nn
	\Gamma_{0}(x,y) &=& g_{\mu_1\mu_2}g_{\nu_1\nu_2}/2 \,,\\\nn
	\Gamma_{1}(x,y) &=& (x-y)_{\mu_1} (x-y)_{\mu_2} g_{\nu_1\nu_2}(d+4)^{-1}\,,\\\nn
	\Gamma_{2}(x,y) &=& (x_{\mu_1} y_{\mu_2}-y_{\mu_1} x_{\mu_2})g_{\nu_1\nu_2}\,,\\\nn
	\Gamma_{3}(x,y) &=& (x_{\mu_1} y_{\mu_2}+y_{\mu_1} x_{\mu_2})g_{\nu_1\nu_2}\,,\\\label{eq:basisGammaXY}
	\Gamma_{4}(x,y) &=& \Delta\cdot\Gamma_{0}(x,y) \,, ~~\Delta=x^2y^2-(xy)^2\,,\\\nn
	\Gamma_{5}(x,y) &=& (  
	x^2y_{\mu_1} y_{\mu_2}+
	y^2x_{\mu_1} x_{\mu_2}-\\\nn
	&&~~
	xy(x_{\mu_1} y_{\mu_2}+y_{\mu_1} x_{\mu_2})
	)g_{\nu_1\nu_2}\,,\\\nn
	\Gamma_{6}(x,y) &=& x_{\mu_1}y_{\nu_1}x_{\mu_2}y_{\nu_2}\,,\\\nn
	\Gamma_{7}(x,y) &=& (y^2)^2 x_{\mu_1} x_{\mu_2}g_{\nu_1\nu_2}\,,~ \\\nn
	\Gamma_{8}(x,y) &=& (x^2)^2 y_{\mu_1} y_{\mu_2}g_{\nu_1\nu_2}\,.
\end{eqnarray}

The tensor basis, Eq.~(\ref{eq:basisGammaXY}), is sufficient in all orders of expansion.
It has been constructed to have simple expansions for the scalar functions $M_i(x,y)$.
The properties of the introduced tensors $\Gamma_{n}$ are collected in Table~\ref{tab:condGdim6tensorProperties}. 
The table includes the leading dimension
of the corresponding condensate  and the specification of which of the tensors contributes  when one of the gluon coordinates coincides with the gauge fixing point, $x=0$ or $y=0$, and when the coordinates $x$ and $y$ lie on one line with the gauge fixing point $x=t y$ (collinearity condition).
The three tensors $\Gamma_{0}$, $\Gamma_{1}$ and $\Gamma_{6}$ are the same
as in~\cite{Grozin:1994hd}.
At dimension-6 order, the tensor $\Gamma_{2}$ is the addition to the $\Gamma_{0}$, $\Gamma_{1}$ tensors considered in~\cite{Mikhailov:1992ug}.
The tensor $\Gamma_{4}$ is introduced to separate the  $\Delta$ term. 
The remaining tensors $\Gamma_{3}$ and $\Gamma_{5}$ are chosen 
so  
that an expansion of the corresponding scalar functions start from dimension-8 condensates.
The special form of the tensor $\Gamma_{5}$ comes from the gluon field strength tensor exchange symmetry in the condensate.

The expansions for the introduced scalar functions are defined in the following way:
\begin{widetext}
\begin{eqnarray}\label{eq:M20}
	M_{0}(x,y) &=&
	G^4_1
	+
	\frac{(x-y)^2}
	{(d-2)(d+2)}\left(
	k^{6,2}_0+\frac{k^{6,2}_1}{d+4}
	+
	\left( k^{8,2}_{0,1} +\frac{k^{8,2}_{1,1}}{2(d+4)}\right)
	\frac{(x-y)^2}{4!}
	\right)
	+\ldots   \,,
	\\\label{eq:M2i}
	M_{i}(x,y)&=&
	\frac{1}
	{(d-2)(d+2)}\left(
	k^{6,2}_{i}
	+
	\frac{k^{8,2}_{i,1}(x-y)^2+2k^{8,2}_{i,2}xy}{4!}
	\right)
	+\ldots   \,,~\text{for~}i=1,2\,,\\\nn
	M_{3}(x,y)&=&
	\frac{k^{8,2}_{3,1}\,(x-y)^2}{4!(d-2)(d+2)}
	+\ldots\,, \\\nn
	M_{i}(x,y)&=&
	\frac{2 k^{8,2}_{i,1}}
	{4!(d-2)(d+2)(d-3)(d+1)}
	+\ldots   \,,~\text{for~}i=4,5,6\,,
\end{eqnarray}
\end{widetext}
where the dimensions-6 coefficients $k^{6,2}_{i}$ and dimensions-8 coefficients $k^{8,2}_{i,j}$  are given in Table \ref{tab:condGdim6} and 
Table \ref{tab:condGdim8},
respectively. 
The expansion is given for space dimension $d$.
The expansions for $M_{7}(x,y)$, $M_{8}(x,y)$ start from the dimension-10 condensates. 
Due to the symmetry concerning gluon field strength tensor exchange, the following scalar functions are related $M_{8}(x,y)=M_{7}(y,x)$.

\begin{table}[h!]
	\caption{\label{tab:condGdim6}
		The dimension-6 coefficients $k^{6,2}_i$ of two-gluon scalar function expansions, Eqs.~(\ref{eq:M20}) and (\ref{eq:M2i}),
		are given in the second column with the subscripts $i$ defined in the first column.
		The third column presents the coefficients in the case of the space dimension $d=4$.
	}
	\begin{center}
		{
			\renewcommand{\arraystretch}{1.3}
			\begin{tabular}{ccc}\hline\hline
				{  ~~~$i$~~~~} & 
				{$k^{6,2}_i$} & 
				{ ~~$d=4$~~}  
				\\\hline	        
				$ 0 $ & 
				$ -(d+2) G_1^6 $ & 
				$ -6 G_1^6 $ 
				\\
				$ 1 $ & 
				$ (d+4) \left[-(d-4) G_1^6-(d-2) G_2^6\right] $ & 
				$ -16 G_2^6 $ 
				\\
				$ 2 $ & 
				$ -(d+2) G_1^6 $ & 
				$ -6 G_1^6 $ 
				\\\hline\hline
			\end{tabular}
		}
	\end{center}
\end{table}

\begin{table*}[t!]
	\caption{\label{tab:condGdim8}
		Expressions for the dimension-8 coefficients $k^{8,2}_{i,j}$ of two-gluon condensate expansions are given in the second column with the  subscripts $i\,,j$ given in the first column. In the third column, the coefficients in the case of gluodynamics in space dimension $d=4$ are given. The last column presents the coefficients when the vacuum gluon field strength tensor is
		(anti-)selfdual $\tilde G^a_{\mu\nu}\equiv i\epsilon_{\mu\nu\al\be}G^{a\al\be}/2=\pm G^a_{\mu\nu}$. 
	}
	\begin{center}
		{
			\renewcommand{\arraystretch}{1.5}
			\begin{tabular}{cccc}\hline\hline
				{~~~$i\,,j$~~~} & 
				{$k^{8,2}_{i,j}$} & 
				{~~~$d=4,~J=0$~~~}  &  
				{~~~Selfdual~~~}	
				\\\hline
				0,1 & 
				$ 2 G^8_{12}+11 G^8_{34}-3 G_{56}^8 $ & 
				$ 2 G^8_{12}+11 G^8_{34} $ & 
				$ 15 G^8_{34} $ 
				\\
				1,1 & 
				$ 2 \left((d-6) G^8_{12}+(13 d-48) G^8_{34}-6 (d-3) G_{56}^8-(d-2) G_{67}^8\right) $ & 
				$ -4 \left(G^8_{12}-2 G^8_{34}\right) $ & 
				$ 0 $ 
				\\
				1,2 & 
				$ (d+4) \left(G^8_{12}-2 G^8_{34}+G_{56}^8\right) $ & 
				$ 8 \left(G^8_{12}-2 G^8_{34}\right) $ & 
				$ 0 $ 
				\\
				2,1 & 
				$ 3 G^8_{12}+24 G^8_{34}-7 G_{56}^8 $ & 
				$ 3 \left(G^8_{12}+8 G^8_{34}\right) $ & 
				$ 30 G^8_{34} $ 
				\\
				2,2 & 
				$ 3 G^8_{34}-G_{56}^8 $ & 
				$ 3 G^8_{34} $ & 
				$ 3 G^8_{34} $ 
				\\
				3,1 & 
				$ -G^8_{12}+2 G^8_{34}-G_{56}^8 $ & 
				$ 2 G^8_{34}-G^8_{12} $ & 
				$ 0 $ 
				\\
				4,1 & 
				$ 2 (d-3) (d-2) G_6^8+(7-3 d) d G^8_{12}+4 (d+3) G^8_{34} $ & 
				$ -4 \left(5 G^8_{12}-7 G^8_{34}\right) $ & 
				$ -12 G^8_{34} $ 
				\\
				5,1 & 
				$ 2 (d-3) (d-2) G_6^8-((d-6) d+3) G^8_{12}-(d (d+6)-3) G^8_{34} $ & 
				$ 5 G^8_{12}-37 G^8_{34} $ & 
				$ -27 G^8_{34} $ 
				\\
				6,1 & 
				$ 6 \left((d-1)^2 G^8_{34}+(d-3) (d-2) G_6^8+(d-4) G^8_{12}\right) $ & 
				$ 54 G^8_{34} $ & 
				$ 54 G^8_{34} $ 
				\\\hline\hline
			\end{tabular}
		}
	\end{center}
\end{table*}

Note that the two-gluon NLC expansion violates translational invariance in the FS gauge, since the expansion depends on the coordinate of the gauge fixing point.
The contribution of the tensor $\Gamma_{2}$ starts in dimension-6 order that is the leading order (LO) where violation of translational invariance occurs. 
The translational invariance of correlators is restored when all contributions to the coefficient function of a given dimension are taken into account~\cite{Mikhailov:1992ug,Nikolaev:1982rq}.

The main result of Section~\ref{sec:nlc} is NLC expansions. The expansions are defined by the Lorentz tensors and scalar functions. The latter are presented in the form of expansion whose  coefficients are linear combinations of the local condensates. For clarity of expansions, we accumulated the coefficients in Tables~\ref{tab:condGdim6}, \ref{tab:condGdim8}, \ref{tab:condGdim8for3}, \ref{tab:condGdim8for4}.

\subsection{Three-gluon NLC}
\label{sec:3Gluon.cond}
The result of the three-gluon nonlocal condensate expansion
up to dimension-8 can be presented by the rank-6 Lorentz tensor and depends on three coordinates: 
\begin{eqnarray}\nn
	&&\!\!\!\!\!\!i\va{\Tr 
		G_{\mu_1\nu_1}\!(x_1)[x_1;x_2]
		G_{\mu_2\nu_2}\!(x_2)[x_2;x_3]
		G_{\mu_3\nu_3}\!(x_3)[x_3;x_1]}
	\\\nn&& =
	\frac{\Gamma(d-2)}{\Gamma(d+1)}\!\!
	\left(\!\prod\limits_{j}^{3} \mathbb{A}({\mu_j,\nu_j})\!\right)
	\frac 12 
	\sum\limits_{i=0}^{7}
	\mathbb{A}
	\Gamma_{i}^{(abc)} M_{i}(x_a,x_b,x_c)
	\\\label{eq:GGGnlc}&& 
	~~~+\ldots\,,
\end{eqnarray}
where the index $(abc)$ could be one of the six permutations of $(123)$ and is used to denote permutations of three sets. 
Each of the sets includes two Lorentz indices and one coordinate: 
($\mu_1$, $\nu_1$, $x_1$), 
($\mu_2$, $\nu_2$, $x_2$), and
($\mu_3$, $\nu_3$, $x_3$).
The scalar functions are denoted by $M_{i}$. The dependence of the rank-6 master Lorentz tensor $\Gamma_{i}^{(abc)}$ on three coordinates $x_1$, $x_2$, $x_3$ is implied:
\begin{eqnarray}\nn
	\Gamma_{i}^{(abc)} M_{i}(x_a,x_b,x_c)=T_{\mu_a\nu_a\mu_b\nu_b\mu_c\nu_c}(x_a,x_b,x_c)=T_{(abc)}\,,
\end{eqnarray}
where $T_{(abc)}$  is short for the rank-6 tensor.
The operator $\mathbb{A}$ is introduced to shorten the expression and restore asymmetry of the condensate with respect to permutations of the gluon field strength tensors.
The operator is defined by the following anti-symmetrization  with respect to permutations
of three sets:
\begin{eqnarray}\nn
	\mathbb{A} T_{(abc)} &=& 
	T_{(123)} - T_{(132)}+
	T_{(231)} - T_{(213)}\\\nn
	&&+T_{(312)} - T_{(321)}\,,
\end{eqnarray}
To shorten the expression, we also define the operator of symmetrization $\mathbb{S}$:
\begin{eqnarray}\nn
	\mathbb{S} T_{(abc)} &=&
	T_{(123)} + T_{(132)}+
	T_{(231)} + T_{(213)}\\\nn
	&&+
	T_{(312)} + T_{(321)}\,,
\end{eqnarray}
Using the operators $\mathbb{S}$ and $\mathbb{A}$ makes the expressions eye-readable and simplifies calculations of OPE. 
The master tensors $\Gamma_{0}^{(abc)}$ are given as follows:
\footnote{
	Note, the symmetrization $\mathbb{S}$ and antisymmetrization  $\mathbb{A}$ act on the product of scalar functions and tensors $\Gamma^{(abc)}_{i}M_{i}(x_a,x_b,x_c)$ in Eq. (\ref{eq:GGGnlc}). While in definitions for the tensors $(\mathbb{A}r_A^{(abc)})$, $(\mathbb{S}r_A^{(abc)})$, and $(\mathbb{S}r_B^{(abc)})$, the operators act only on the tensor in the parentheses.}
\begin{eqnarray}\nn
	\Gamma_{0}^{(abc)} &=& g_{\nu_c\mu_a}g_{\nu_a\mu_b}g_{\nu_b\mu_c}/3 \,,\\\nn
	\Gamma_{1}^{(abc)} &=& 
	\frac{1}{d+1}\!
	\left[(\mathbb{S}r_A^{(abc)})\!+\!(\mathbb{S}r_B^{(abc)}/2)\right]
	{(x_c)}_{\rho} {(x_a)}_{\si} \,,\\\nn
	\Gamma_{2}^{(abc)} &=& \left[{(x_c)}_{\rho} {(x_a)}_{\si}\!-\!{(x_c)}_{\si} {(x_a)}_{\rho}\right]r_A^{(abc)}\,,\\\nn
	\Gamma_{3}^{(abc)} &=& {(x_c)}_{\rho} {(x_a)}_{\si}r_B^{(abc)} \,,\\\label{eq:tensorsrArB} 
	\Gamma_{4}^{(abc)} &=& \left[(\mathbb{A}r_A^{(abc)})\!-\!(\mathbb{S}r_B^{(abc)}/2)\right]{(x_c)}_{\rho} {(x_a)}_{\si} \,,\\\nn
	\Gamma_{5}^{(abc)} &=&
	\left[{(x_c)}_{\rho} {(x_a)}_{\si}\!+\!{(x_c)}_{\si} {(x_a)}_{\rho}\right]r_A^{(abc)} \,,\\\nn
	\Gamma_{6}^{(abc)} &=& {(x_b)}_{\rho} {(x_b)}_{\si}r_A^{(abc)}\,,\\\nn
	\Gamma_{7}^{(abc)} &=& \left[
	{(x_a)}_{\rho} {(x_a)}_{\si}\!+\!
	{(x_b)}_{\rho} {(x_b)}_{\si}\!+\!
	{(x_c)}_{\rho} {(x_c)}_{\si}\right] r_A^{(abc)}\,,\\\nn
	r_A^{(abc)}&=&	
	g_{\nu_c\rho}
	g_{\si\mu_a}
	g_{\nu_a\mu_b}
	g_{\nu_b\mu_c}\,,~~
	\\\nn 
	r_B^{(abc)} &=&
	g_{\mu_a\mu_c}
	g_{\nu_a\nu_c}
	g_{\mu_b\rho}
	g_{\nu_b\si}\,.
\end{eqnarray}
The choice for the tensors has been motivated by the simplicity of the scalar function expansions considered up to dimension-8 order: 
\begin{eqnarray}\nn
	M_{0}(x,y,z)&=&
	G^6_1
	+
	\frac{k^{8,3}_0}
	{8 (d-3)(d+2)}\cdot\\\nn
	&&
	\cdot
	\frac{(y-x)^2+(x-z)^2+(z-y)^2}4
	+\ldots   \,,\\\nn
	M_{i}(x,y,z)&=&
	\frac{k^{8,3}_i}
	{8(d-3)(d+2)}
	+\ldots   \,,~~\text{for}~i\geq 1\,,
\end{eqnarray}
where the coefficients $k^{8,3}_i$ are collected in Table \ref{tab:condGdim8for3}.
The same as in the two-gluon case, the gauge link path is a broken line with apexes at the gauge fixing point; therefore, the link can be omitted $[x;y]=[x;0][0;y]=1$ due to the gauge condition, Eq. (\ref{eq:FPG}).
Note that only the C-even part contributes in Eq.~(\ref{eq:GGGnlc}): \begin{eqnarray}\nn
	&&\va{\Tr G_{\mu_1\nu_1}(x)G_{\mu_2\nu_2}(y)G_{\mu_3\nu_3}(z)}
	\\\nn
	&&~~~~~~=
	\frac 12\va{\Tr G_{\mu_1\nu_1}(x)\left[G_{\mu_2\nu_2}(y),G_{\mu_3\nu_3}(z)\right]}
\end{eqnarray}
while the C-odd contribution is equal to zero
\begin{eqnarray}\nn
	\va{\Tr G_{\mu_1\nu_1}(x)
		\left\{G_{\mu_2\nu_2}(y),G_{\mu_3\nu_3}(z)\right\}}=0\,,
\end{eqnarray}
which is explicitly confirmed by the obtained expansion, Eq. (\ref{eq:GGGnlc}), up to dimension-8 order.
The C-parity causes antisymmetrization of the condensate with respect to permutations of the gluon field strength tensors in three-gluon NLC.

\begin{table*}[t]
	\caption{\label{tab:condGdim8for3}
		Expressions for the dimension-8 coefficients $k^{8,3}_{i}$ of the three-gluon condensate expansions are given in the second column with the subscript $i$ given in the first column.  In the third column, the coefficients in the case of gluodynamics in space dimension $d=4$ are given. The last column presents the coefficients when the vacuum gluon field strength tensor is
		(anti-)selfdual $\tilde G^a_{\mu\nu}\equiv i\epsilon_{\mu\nu\al\be}G^{a\al\be}/2=\pm G^a_{\mu\nu}$. 
	}
	\begin{center}
		{
			\renewcommand{\arraystretch}{1.5}
			\begin{tabular}{cccc}\hline\hline
				{~~~$i$~~~} & 
				{$k^{8,3}_{i}$} & 
				{~~~$d=4,~J=0$~~~}  &  
				{~~~Selfdual~~~}	
				\\\hline
				$ 0 $ & 
				$ 4\left(G^8_{12}-4 (d-2) G^8_{34}+(d-3) G_{56}^8\right) $ & 
				$ 4\left(G^8_{12}-8 G^8_{34}\right) $ & 
				$ -24 G^8_{34} $ 
				\\
				$ 1 $ & 
				$ -2 (d-3) (d-2) G_6^8+(d-4) (d-1) G^8_{12}+4 (d-1) G^8_{34} $ & 
				$ 12 G^8_{34} $ & 
				$ 12 G^8_{34} $ 
				\\
				$ 2 $ & 
				$ 4 G^8_{12}-4 (d+1) G^8_{34} $ & 
				$ 4 \left(G^8_{12}-5 G^8_{34}\right) $ & 
				$ -12 G^8_{34} $ 
				\\
				$ 3 $ & 
				$ 2 (d-2) G^8_{12}-8 G^8_{34} $ & 
				$ 4 \left(G^8_{12}-2 G^8_{34}\right) $ & 
				$ 0 $ 
				\\
				$ 4 $ & 
				$ (d-2) G^8_{12}-4 G^8_{34} $ & 
				$ 2 \left(G^8_{12}-2 G^8_{34}\right) $ & 
				$ 0 $ 
				\\
				$ 5 $ & 
				$ -2 (d-3) \left(G^8_{12}-2 G^8_{34}+G_{56}^8\right) $ & 
				$ -2 \left(G^8_{12}-2 G^8_{34}\right) $ & 
				$ 0 $ 
				\\
				$ 6 $ & 
				$ -2 (d-3) \left(G^8_{12}-2 G^8_{34}+G_{56}^8\right) $ & 
				$ -2 \left(G^8_{12}-2 G^8_{34}\right) $ & 
				$ 0 $ 
				\\
				$ 7 $ & 
				$ -2 \left(2 (d-5) G^8_{34}-(d-3) G_{56}^8+G^8_{12}\right) $ & 
				$ -2 \left(G^8_{12}-2 G^8_{34}\right) $ & 
				$ 0 $ 
				\\\hline\hline
			\end{tabular}
		}
	\end{center}
\end{table*}

\begin{table*}[t!]
	\caption{\label{tab:condGdim8for4}
		Expressions for the dimension-8 coefficients $k^{8,4}_{i}$ of the four-gluon condensate expansion are given in the second column with the subscript $i$ given in the first column. 
		In the third column, the coefficients in the case of gluodynamics in space dimension $d=4$ are given. The last column presents the coefficients when the vacuum gluon field strength tensor is
		(anti-)selfdual $\tilde G^a_{\mu\nu}\equiv i\epsilon_{\mu\nu\al\be}G^{a\al\be}/2=\pm G^a_{\mu\nu}$. 
	}
	\begin{center}
		{\renewcommand{\arraystretch}{1.5}
			\begin{tabular}{cccc}\hline\hline
				{~~~$i$~~~} & 
				{$k^{8,4}_{i}$} & 
				{~~~~~~~~~~~~~~~$d=4,~J=0$~~~~~~~~~~~~~~~}  &  
				{~~~~~~~~~Selfdual~~~~~~~~~}	
				\\\hline    
				$ 1 $ & 
				$ (d+1) \left[(d+1) G^8_{34}-G^8_{12}\right] $ & 
				$ -5 \left(G^8_{12}-5 G^8_{34}\right) $ & 
				$ 15 G^8_{34} $ 
				\\
				$ 2 $ & 
				$ \left(d^2+3\right) G_4^8+(1-d) G_1^8-d G_2^8+(1-d) G_3^8 $ & 
				$ -3 G_1^8-4 G_2^8-3 G_3^8+19 G_4^8 $ & 
				$ -3 \left(G_1^8+G^8_{34}\right) $ 
				\\
				$ 3 $ & 
				$ -(d+1) \left[(d-2) G^8_{12}-4 G^8_{34}\right] $ & 
				$ -10 \left(G^8_{12}-2 G^8_{34}\right) $ & 
				$ 0 $ 
				\\
				$ 4 $ & 
				$ \left(d^2-d+2\right) G_1^8-4 d G_3^8-4 (d-1) G_4^8+2 G_2^8 $ & 
				$ 2 \left(7 G_1^8+G_2^8-8 G_3^8-6 G_4^8\right) $ & 
				$ 9 G_1^8-6 G^8_{34} $
				\\\hline\hline
			\end{tabular}
		}
	\end{center}
\end{table*}

\subsection{Four-gluon NLC}
\label{sec:4Gluon.cond}
We present the four-gluon condensate in the following 
form~\footnote{The gauge links are omitted here $[x_i;x_j]=[x_i;0][0;x_j]=1$ 
	for the gauge condition, Eq. (\ref{eq:FPG}), and their paths are chosen to be a broken line with apexes at $0$ and end points $x_i$ and $x_j$.}:
\begin{eqnarray}\nn
&& \va{\Tr(G_{\mu_1\nu_1}(x_1)G_{\mu_2\nu_2}(x_2)G_{\mu_3\nu_3}(x_3)G_{\mu_4\nu_4}(x_4))}
	\\\label{eq:GGGGttttexpand}
&&\hskip 10mm =
	\frac{\Gamma(d-3)}{\Gamma(d+3)}
	\left(\prod\limits_{n=1}^{4} \mathbb{A}({\mu_n,\nu_n})\right)
	\\\nn
&&\hskip 12mm \times
	\sum\limits_{m=1}^{4}
	\Gamma_{m}
	M_{m}(x_1,x_2,x_3,x_4)
	+\ldots \,,
\end{eqnarray}
where $M_{m}$ are the scalar functions and $\Gamma_{m}$ are the Lorentz tensors that contribute starting with the leading dimension-8 order.
The four-gluon condensate is symmetric with respect to cyclic permutations of the gluon field strength tensors and the reflection of their order in the trace.
To respect these symmetries, the scalar functions $M_{m}$ should be invariant 
to cyclic permutations of the arguments $(x_1,x_2,x_3,x_4)$ and the reflection of their order:
\begin{eqnarray}\nn 
	&&M_{m}(x_1,x_2,x_3,x_4) = 	k^{8,4}_{m} +
	a_m (x_1^2+x_2^2+x_3^2+x_4^2) 
	\\\nn &&~~~~~~
	+ b_m (x_1\cdot x_2+x_2\cdot x_3+x_3\cdot x_4+x_4\cdot x_1)  
	\\\nn && ~~~~~~
	+ c_m (x_1\cdot x_3+x_2\cdot x_4) +\cdots\,.
\end{eqnarray}
The leading order terms $k^{8,4}_{m}$  are given in   Table~\ref{tab:condGdim8for4} in terms of the dimension-8 local condensates. 
The dimension-10 order coefficients $a_m$, $b_m$, and $c_m$ are not considered here, as we work in dimension-8 order.
The Lorentz tensors respect the discussed symmetry due to antisymmetrization of four pairs of indices $\mu_n\nu_n$, $n=1,2,3,4$:
\begin{eqnarray}\nn
	\Gamma_{1} &=& r^{(1234)}_A\,,~	
	\Gamma_{2}  =  r_A^{(1234)}+r_A^{(2314)}+r_A^{(3124)}\,,\\\nn
	\Gamma_{3} &=& r^{(1234)}_B\,,~
	\Gamma_{4}  = r_B^{(1234)}+r_B^{(2314)}+r_B^{(3124)}\,,\\
	\label{eq:tensorABi}
	r_A^{(ijkl)} &=& 	
	g_{\nu_k\mu_l}
	g_{\nu_l\mu_i}
	g_{\nu_i\mu_j}
	g_{\nu_j\mu_k}\,,\\\nn
	r_B^{(ijkl)}  &=&	g_{\mu_i\mu_k} g_{\nu_i\nu_k}	g_{\mu_j\mu_l}	g_{\nu_j\nu_l}	/4\,.
\end{eqnarray}
The considered vacuum expectation corresponds to the general case
$\va{G^{a_1}_{\mu_1\nu_1}G^{a_2}_{\mu_2\nu_2}G^{a_3}_{\mu_3\nu_3}G^{a_4}_{\mu_4\nu_4}}$ condensate that is contracted with the color tensor $\Tr t^{a_1}t^{a_2}t^{a_3}t^{a_4}$, where $a_i$ are the color labels (i.e. group index) of the gluon field strength tensors. 
This expansion can be used for calculating OPE of the  quark fields correlators. 
From the obtained expression (\ref{eq:GGGGttttexpand}) we can extract the expansion of the four-gluon condensate $\va{G^{a_1}_{\mu_1\nu_1}G^{a_2}_{\mu_2\nu_2}G^{a_3}_{\mu_3\nu_3}G^{a_4}_{\mu_4\nu_4}}$ contracted with the color tensor $f^{ba_1a_2}f^{ba_3a_4}$ which is useful for OPE calculations related to glueball states:
\begin{eqnarray}\label{eq:GGGGffExpand}
	&&\va{\Tr([G_{\mu_1\nu_1},G_{\mu_2\nu_2}][G_{\mu_3\nu_3},G_{\mu_4\nu_4}])}
	\\\nn && ~~~~~~
	=
	\frac{\Gamma(d-3)}{\Gamma(d+3)}
	\left(\prod\limits_{j=1}^{4} \mathbb{A}(\mu_j,\nu_j)\right)
	\sum\limits_{i=1}^{2}
	\tilde \Gamma_{i} \tilde M_i
	\,,\\\nn && 
	\tilde \Gamma_1 = (r^{(1234)}_A-r^{(3124)}_A)\,,~~~
	\tilde \Gamma_2  = (r^{(1234)}_B-r^{(3124)}_B)\,,
		\\\nn &&
	\tilde M_1=2k^{8,4}_1\,,~~~ \tilde M_2=2k^{8,4}_3\,,
\end{eqnarray}
where $G_{\mu\nu}\equiv  g G_{\mu\nu}^a(0) t^a$,
the tensors $r^A_i$, $r^B_i$ are defined in (\ref{eq:tensorABi}),
and the coefficients $k^{8,4}_1$ and $k^{8,4}_3$ are given in Table~\ref{tab:condGdim8for4}. 
The expansion (\ref{eq:tensorABi}) coincides with the one given in \cite{Hao:2005hu}.

\section{Usage of gluon NLC expansions}
\label{sec:usage}

In the previous section, we obtained two-, three-, and four-gluon NLC expansions in the FS gauge.
This section is dedicated to a discussion of the practical importance of the obtained expansions.
The usual way of OPE calculations includes applying the Taylor expansion of vacuum fields, see Eq.~(\ref{eq:Gexpansion}). 
In the case of gluon condensates, the Taylor expansion in mass dimension-$D$ order leads to intermediate expressions in the form of the rang-$D$ tensor.
In dimension-8 order of OPE, there are seventy tensor condensates formed by the gluon field strength tensor and its derivatives, e.g.,  $\va{\Tr G_{\mu_1\mu_2} D_{\mu_3}G_{\mu_4\mu_5}D_{\mu_6}G_{\mu_7\mu_8}}$. 
The tensor expansion of these condensates leads to an expression defined by eleven scalar local condensates given in Eq.~(\ref{eq:G6}) and Eq.~(\ref{eq:G8}).

Using NLC expansions causes a significant reduction of computational work by jumping over the Taylor and tensor expansions.  
As we will see in the next subsection, the application of NLC expansions is especially efficient in OPE of a vacuum correlator for the currents with the gluon field strength tensor, such as currents of glueball and hybrid states~\cite{
	Mathieu:2009sg,Amato:2015ipe,Cho:2015rsa,He:2015owa,
	Gutsche:2016wix,Zhang:2016vcx,Azizi:2017xyx,Csorgo:2018uyp,Xu:2018cor,Gastaldi:2018ztu,Souza:2019ylx,Ryttov:2019aux,Khlebtsov:2020rte,Zhang:2021itx,Rinaldi:2022dyh,Ballon-Bayona:2017sxa,Kaptari:2019ghz,Kaptari:2020qlt,Llanes-Estrada:2021evz}.

\begin{figure*}[t]	
	\centerline{		
		\begin{overpic}[width=0.97\textwidth 
			]{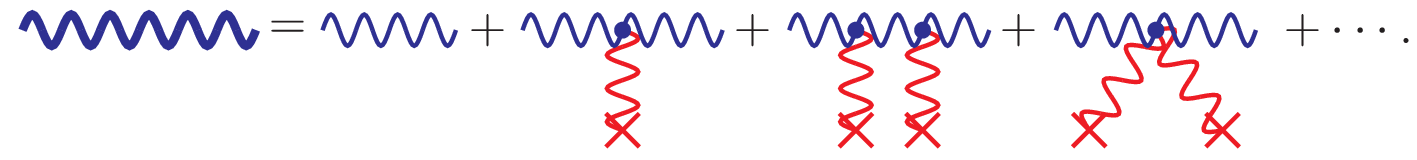}
			\put (10,60) {
				$D$\hspace{0.21\textwidth}
				$D_0$\hspace{0.13\textwidth}
				$D_1$\hspace{0.15\textwidth}
				$D_{21}$\hspace{0.15\textwidth}$D_{22}$}
		\end{overpic}
	}
	\caption{ 
		\label{fig:background-propagator}
		Contributions to the gluon propagator in the external fields,
		Eq.~(\ref{eq:D4}). 
	}
\end{figure*}

In the QCD SR approach, the correlator OPE serves as a source of information on hadron parameters.
The OPE of vacuum correlator $\Pi$ based on the dimension-$D$ truncated NLC expansions
is the same as OPE defined by the local condensates up to dimension-$D$ order:
\begin{equation}\label{eq:local2NLC}
	\Pi	=\Pi_\text{pert}+\sum\limits_{i=1}^{N_D}\Pi_{\text{NLC-}i}
	=\Pi_\text{pert}+\sum\limits_{j=3}^D\Pi_j\,,
\end{equation}
where $\Pi_\text{pert}$ is the perturbative contribution, $\Pi_j$ is the nonperturbative contribution
\footnote{
	The minimum dimension condensate is the quark condensate
	$\va{\bar qq}$ whose  mass dimension is three; therefore, summation of $\Pi_j$ starts from $j=3$ in the general case.}
of dimension-$j$, and $\Pi_{\text{NLC}-i}$ is the $i$-th NLC contribution.
The number of NLC-based diagrams that contribute to dimension-D is denoted by $N_D$. 
The equation~(\ref{eq:local2NLC}) reflects the idea of the OPE rearrangement in terms of NLCs.

The NLC expansion can be widely applied due to its universality. 
The usage of NLC expansions is demonstrated by the vacuum correlators of two-gluon and three-gluon glueball currents with quantum numbers $0^{\pm+}$. 
The discussion below is given in general terms, while the technical details
are placed in the Appendices.
In particular, the full set of Feynman rules needed for such calculations is presented in Appendix~\ref{sec:subNLCaplication}, while 
Appendix~\ref{sec:subDia2} provides the detailed calculation
of one of the contributions to OPE of the glueball current correlators.

Before considering these applications,
we want to refresh some aspects of OPE calculations~\cite{Novikov:1983gd,Grozin:1994hd}
in the next subsection.
We cover aspects of the background approach~\cite{DeWitt:1967uc,tHooft:1976snw,Huang:1989gv} 
related to glueball studies
and the gluon propagator and its precalculated expansions in the background gluon field.

\subsection{Background field approach}

In the background field approach, the total gluon field is considered as  a compound of two fields
$\bar A^a_\mu=A^a_\mu+a^a_\mu$:
the perturbative quantum gluon field $a^a_\mu$ and background field $A^a_\mu$.
To keep the gluon propagator of the quantum field $a^a_\mu$ in Feynman gauge form, one should add the generalization gauge fixing term $(D_\mu a^a_\mu)^2/(-2)$ to the Lagrangian
that causes modification of the interaction between background and quantum fields. 
In Fig. \ref{fig:backgroundRules}, we provide the Feynman rules for vertices of quantum gluon field interaction with background fields
for the case of three- and four-gluon vertices where two fields are quantum and the rest are background fields.
Using these rules, one obtains expansion of the gluon propagator $D$ in the external field up to dimension-4 order:
\begin{eqnarray}\nn
	-iD(p)
	&=& -i D_0+(-i)^2D_0V_1D_0
	\\\nn
	&&+\frac{2}{2!}(-i)^3D_0V_1D_0V_1D_0
	\\\nn
	&&+(-i)^2D_0V_2D_0+\ldots~,
\end{eqnarray}
where $D_0$ is the free propagator, $V_1$ and $V_2$ are the vertices of interaction between quantum field and  background field. Definitions and graphical notations of vertices are given in Fig.~\ref{fig:backgroundRules}. 
The gluon propagator can be expressed by four terms given in 
graphical form in Fig. \ref{fig:background-propagator}:
\begin{eqnarray}\label{eq:D4}
	D_{\al\be}(p)&=&i\int d^4x e^{ipx} \va{T\{a_\al(x)a_\be(0)\}}
	\\\nn
	&=& 
	D_{0\al\be}+D_{1\al\be}+D_{21\al\be}+D_{22\al\be}+\ldots\,.
\end{eqnarray}
Note that there is an additional diagram for the third term $D_{21}$ that gives combinatoric factor 2.
The gluon propagator expansion was obtained in~\cite{Shuryak:1981pi} (see also~\cite{Novikov:1983gd,Grozin:1994hd}).
For reader's convenience, we present the result for each term
\begin{eqnarray}\nn
	D_{0\al\be}&=&\frac1{p^2}\de_{\al\be}\,,\\
	D_{1\al\be}&=& \frac{2G_{\al\be}}{p^4}
	+\frac1{p^6}\left(\frac23ip_\mu J_\mu\de_{\al\be}
	+4ip_\la D_\la G_{\al\be}\right)
	\nonumber\\\nn
	&&+\frac1{p^8}\Bigg[-2p_\la D_\la p_\mu J_\mu\de_{\al\be}
	\Bigg.\\\nn && \Bigg.
	-2\left(4(p_\la D_\la)^2-p^2D^2\right)G_{\al\be}
	\Bigg]+\Delta D_{1\al\be}\,,
	\nonumber
	\\\nn
	D_{21\al\be}&=& \frac{4}{p^6}G_{\al\la}G_{\la\be}\,,
	\\\nn  
	D_{22\al\be} &=& \frac{\de_{\al\be}}{2p^8}\left(p^2G_{\mu\nu}G_{\mu\nu}
	+4p_\mu G_{\mu\la}G_{\la\nu}p_\nu\right)
	\\ && -\Delta D_{1\al\be}\,,
	\nn
	\\\nn
	\Delta D_{1\al\be}&=& \frac{1}{p^8}\left(p^2[G_{\al\mu},G_{\mu\be}]
	+2p_\mu p_\nu[G_{\al \mu},G_{\be \nu}] \right)\,,
\end{eqnarray}
where we use compact matrix notation for the field strength tensor $G=G^{ac}=g f^{abc}G^b$ and the current $J=J^{ac}=g f^{abc}J^b$.

\begin{figure*}[t]
	\centerline{
		\includegraphics[width=0.24\textwidth]{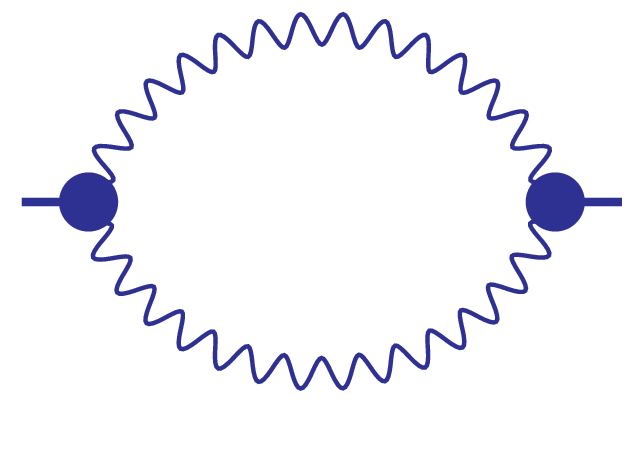}
		\includegraphics[width=0.24\textwidth]{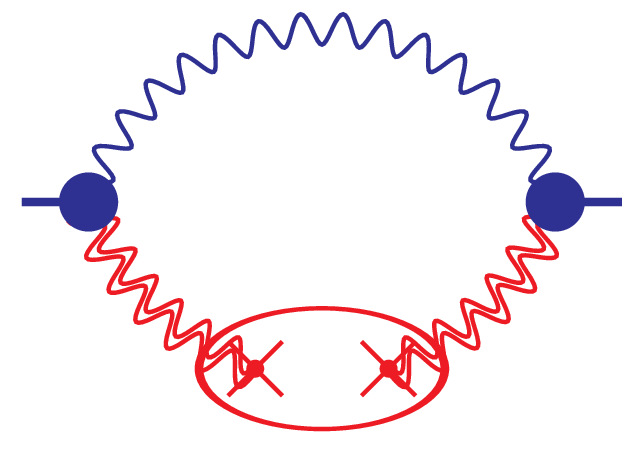}
		\includegraphics[width=0.24\textwidth]{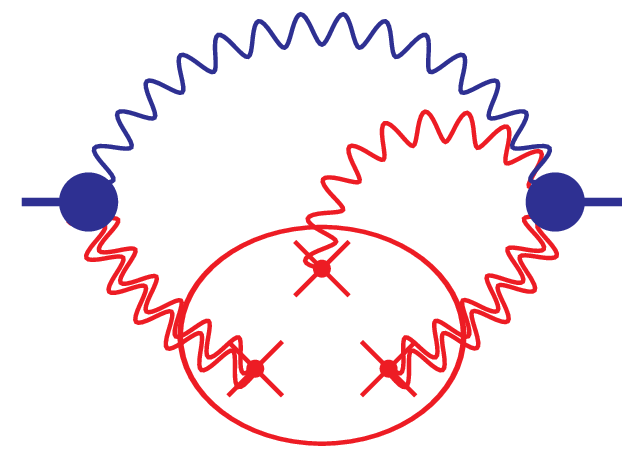}
		\begin{overpic}[width=0.24\textwidth 
			]{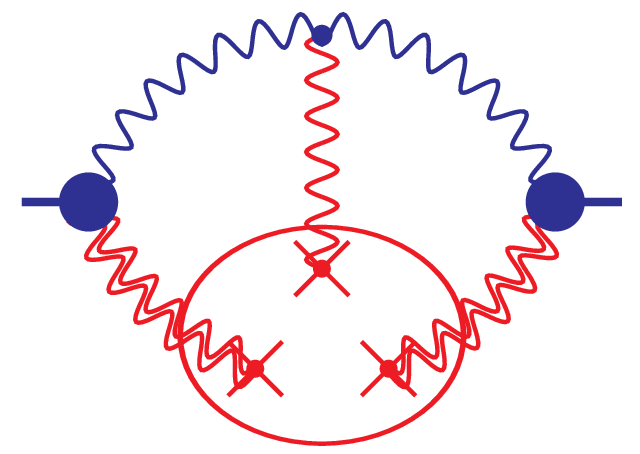}
			\put (-330,-7) {\small{
					(a) $\Pi^\pm_\text{LO}$\hspace{30mm} 
					(b) $\Pi^\pm_{\text{NLC-}0}$\hspace{29mm}
					(c) $\Pi^\pm_{\text{NLC-}1}$\hspace{29mm}
					(d) $\Pi^\pm_{\text{NLC-}2}$}}
		\end{overpic}
	}
	\caption{ 
		\label{fig:diaJJ2gG2G3}
		Diagrams for the LO perturbative term (the first diagram) and nonperturbative contributions to the correlator. The expansions of the depicted terms start from dimension-4 condensate ($\va{G^2}$ for the second diagram) and dimension-6 ($\va{G^3}$ for the third and the fourth diagrams).
		NLC is represented by crosses that denote vacuum gluon fields (single line)
		or vacuum gluon field strength tensors (double line). 
		The Feynman rules for the elements of the diagrams are given in Appendix~\ref{sec:subNLCaplication}. 
	}
\end{figure*}
\begin{figure*}[t]	
	\centerline{
		\includegraphics[width=0.24\textwidth]{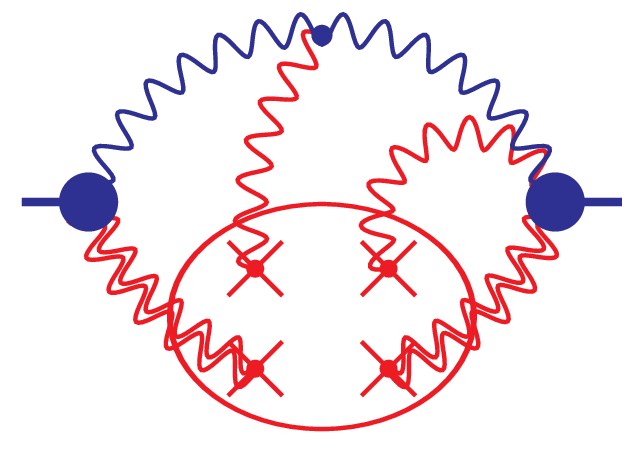}
		\includegraphics[width=0.24\textwidth]{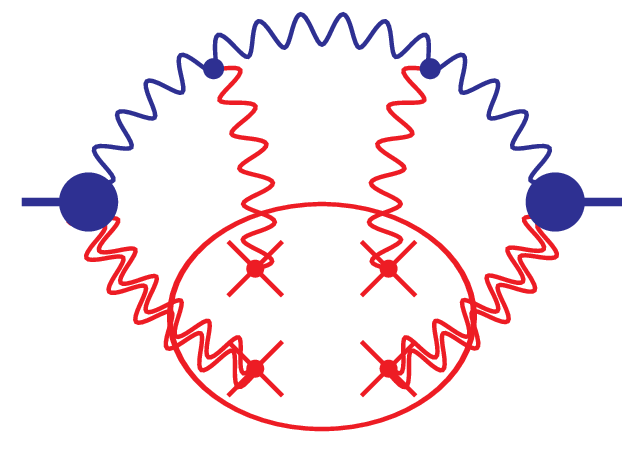}
		\begin{overpic}[width=0.24\textwidth 
			]{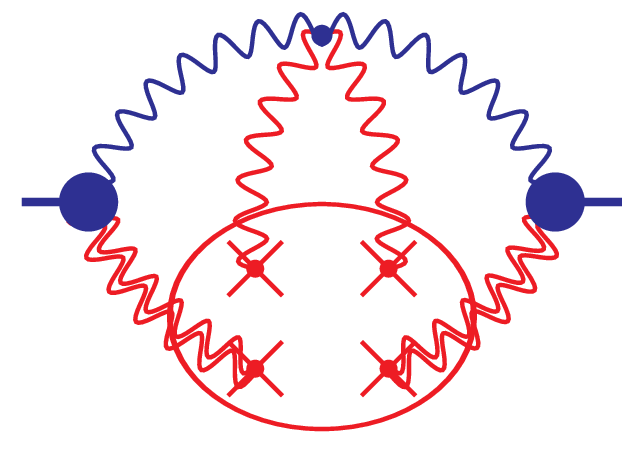}
			\put (-210,-7) {\small{
					(e) $\Pi^\pm_{\text{NLC-}3}$\hspace{29mm}
					(f) $\Pi^\pm_{\text{NLC-}4}$\hspace{29mm}
					(g) $\Pi^\pm_{\text{NLC-}5}$
			}}
		\end{overpic}
	}
	\caption{ 
		\label{fig:diaJJ2gG4}
		Diagrammatic representation of three groups of nonperturbative contributions  to the correlator $\Pi^{P}(q)$ OPE that starts from dimension-8 order. 		The Feynman rules for the elements of the diagrams are given in Appendix~\ref{sec:subNLCaplication}.}
\end{figure*}

The glueball and hybrid state currents include the gluon field strength tensor.
For such currents, the correlator OPE has terms that depend on 
the modified propagator with fields derivatives:
\begin{eqnarray}\nn
	D_{\al\be,\mu\nu}(p)&=&i\int d^4x e^{ipx} \va{T\{\partial_\mu a_\al(x) \partial_\nu a_\be(0)\}}
	\\\nn 
	&=& 
	\frac{g_{\al\be}q_\mu q_\nu}{q^2}+\frac{2q_\mu q_\nu G_{\al\be}}{q^4}
		\\\nn && 
	-\frac{g_{\al\be} q_\mu q_\rho G_{\nu \rho}}{q^4}+\ldots\,.
\end{eqnarray}
The first two terms can be easily obtained from propagator's expansion given in Eq. (\ref{eq:D4}),
while the third term is additional and related to the derivative of the quantum field $a_\be(0)$.
Therefore, in glueball and hybrid-related studies within QCD SR, one needs to calculate additional background field corrections compared to those given in Eq.~(\ref{eq:D4}).

The usage of NLCs truncated series allows the calculation of all background field corrections to the final expression for the correlator OPE without applying the gluon propagator expansion.
This approach to perform OPE appears especially useful for
C-odd glueball studies~\cite{Pimikov:2016pag,Pimikov:2017bkk,Pimikov:2017xap},
where the glueball currents include high derivatives of the gluon field strength tensor. 
The suggested approach can be widely applied due to the universality of NLC expansions. 
In the next subsections, we demonstrate the application of the approach to OPE of glueball current correlators.

\subsection{Two-gluon $0^{\pm +}$ glueballs}

QCD SR for glueballs were first considered in~\cite{Novikov:1979ux,Novikov:1979va} 
by evaluating the correlator
\begin{equation}\label{eq:correlatorP}
	\Pi^{P}(q) = i\int\!\! d^4x\, e^{iqx} \va{T\{ J^{P}_2(0)J^{P\dagger}_2(x)\}}
\end{equation}
of the two-gluon current for the scalar $0^{++}$ and pseudo-scalar $0^{-+}$ 
glueball states:
\begin{equation}\nn
	J^{P}_2(x)=
	\alpha_S
	\delta^{a_1 a_2} T^{P}_{\mu\nu} 
	G^{a_1}_{\mu_1\nu_1}(x)G^{a_2}_{\mu_2\nu_2}(x)\,,
\end{equation}
where Lorentz tensors $T^{+}_{\mu\nu}\equiv g_{\mu_1\mu_2}g_{\nu_1\nu_2}$, and
$T^{-}_{\mu\nu}\equiv i\epsilon_{\mu_1\nu_1\mu_2\nu_2}/2$ specify the parity $P=\pm$.
The OPE of the correlator can be presented as
\begin{equation}\label{eq:correlatorP.terms}
	\Pi^{P}(q) = \Pi^{P}_\text{LO}+\Pi^{P}_4+\Pi^{P}_6+\Pi^{P}_8
	+\cdots\,,
\end{equation}
where $\Pi^{P}_\text{LO}$ is the leading order perturbative contibution,
and
the nonperturbative corrections $\Pi^{P}_n$ include only condensates with even dimension if we do not consider a radiative correction.
This OPE was calculated in~\cite{Novikov:1979ux,Novikov:1979va} 
up to dimension-8 where only gluon condensates were taken into account.
The radiative correction to the dimension-4 and dimension-6 gluon condensate terms 
was obtained in~\cite{Kataev:1981gr,Kataev:1981aw,Bagan:1989vm}.   

We rearrange the contributions to OPE in Eq.~(\ref{eq:correlatorP.terms}), by collecting terms arising from one NLC 
and a specific hard part of the diagram
\begin{equation}\nn
	\Pi^{P}(q) = \Pi^{P}_\text{LO}+\sum\limits_{i=0}^5\Pi^{P}_{\text{NLC-}i}+\cdots\,,
\end{equation}
where $\Pi^{P}_{\text{NLC-}i}$ is one of the six NLC groups of OPE contributions to $\Pi^{P}(q)$ 
that provide contribution to the dimension-8 condensates or lower dimension terms.
Each group $\Pi^{P}_{\text{NLC-}i}$ is depicted by one diagram in Fig. \ref{fig:diaJJ2gG2G3} and Fig. \ref{fig:diaJJ2gG4}. 
The NLCs that have dimensions higher than eight are not considered in our work.
The crosses on the figures specify the background field and the blob around the crosses denotes NLC. 
There are two types of lines: the single line represents gluon fields; the double line depicts the gluon field strength tensor. The red color and the cross at the end of the line denote the soft part that forms a vacuum condensate. 
The blue color of the line denotes the hard part of the diagram. 
For condensates with a gluon fields (single red line with the cross at the end),
we apply Eq. (\ref{eq:FS-integral}) to express the term through obtained NLCs expansions.
The Feynman rules for the  diagrams are given in Appendix~\ref{sec:subNLCaplication}.

\begin{figure*}[t]
	\centerline{
		\includegraphics[width=0.16\textwidth]{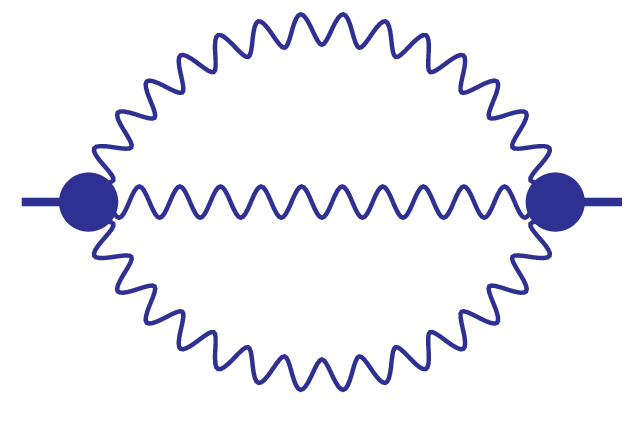}
		\includegraphics[width=0.16\textwidth]{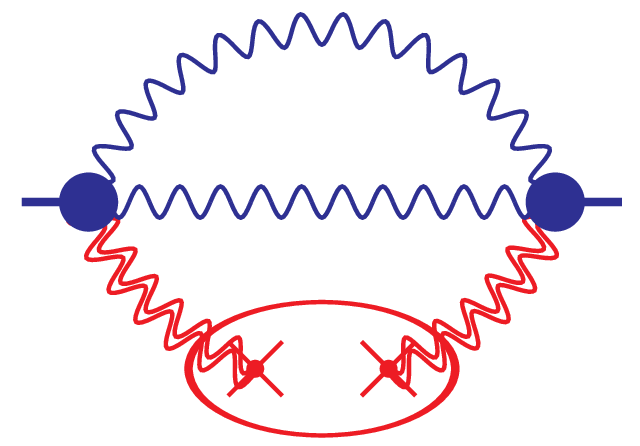}
		\includegraphics[width=0.16\textwidth]{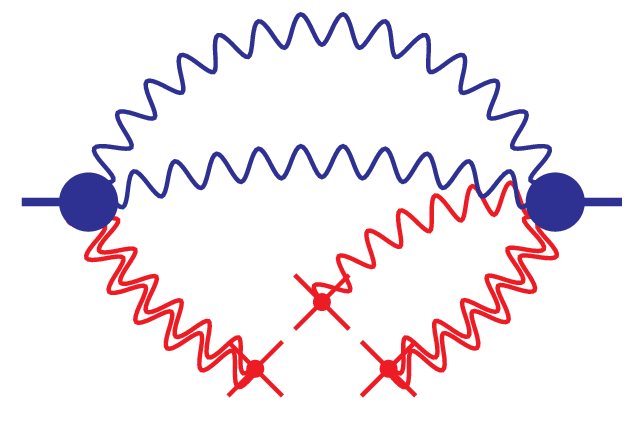}
		\includegraphics[width=0.16\textwidth]{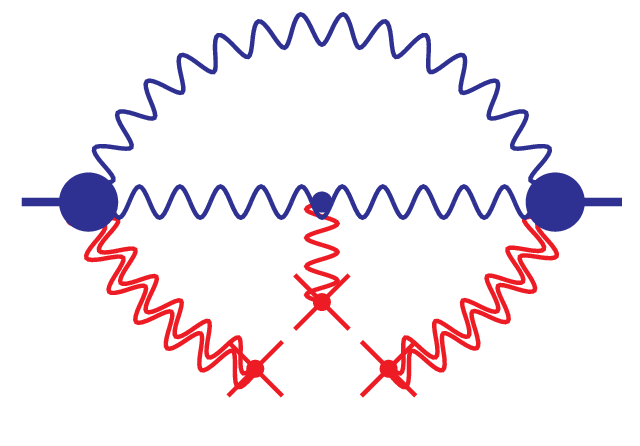}
		\includegraphics[width=0.16\textwidth]{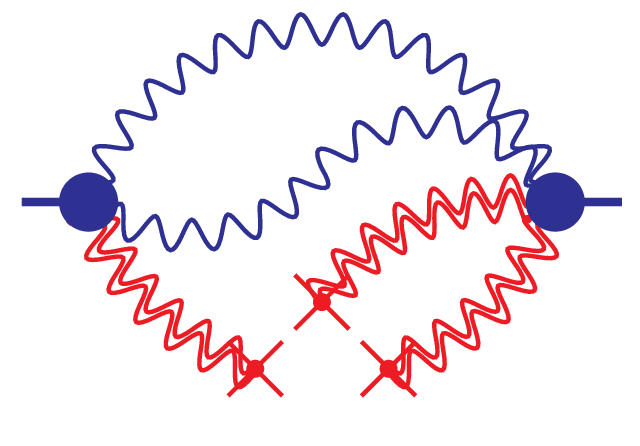}
		\begin{overpic}[width=0.16\textwidth 
			]{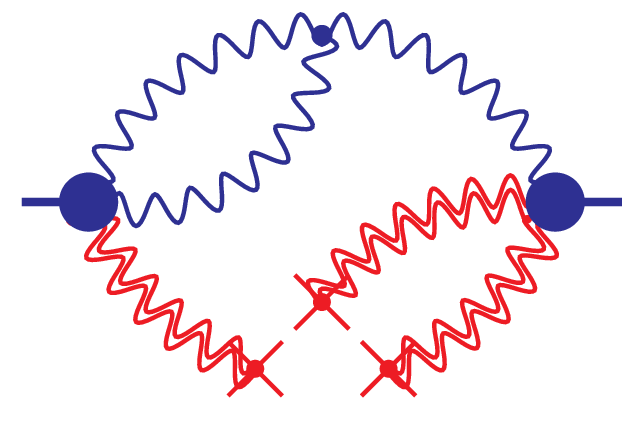}
			\put (-400,-10) {\small{
					(a) $\tilde\Pi^\pm_\text{LO}$ \hspace{0.1\textwidth} 
					(b) $\tilde\Pi^\pm_1$\hspace{0.105\textwidth} 
					(c) $\tilde\Pi^\pm_2$\hspace{0.105\textwidth} 
					(d) $\tilde\Pi^\pm_3$\hspace{0.105\textwidth} 
					(e) $\tilde\Pi^\pm_4$\hspace{0.105\textwidth} 
					(f) $\tilde\Pi^\pm_5$
			}}
		\end{overpic}
	}
	\caption{ 
		\label{fig:diaJJtopG4}
		The diagrams representing the LO perturbative term, dimension-4 and dimension-6  contributions to the correlator OPE of three-gluon currents. 
		Each term is given by $\tilde \Pi_k^\pm$ where $k$ is the number
		given below the diagram.		The Feynman rules for the elements of the diagrams are given in Appendix~\ref{sec:subNLCaplication}.
	}
\end{figure*}

The first diagram in Fig. \ref{fig:diaJJ2gG2G3} is the LO perturbative contribution,
while the second diagram could contribute from dimension-4 to higher dimensions.
The OPE for the third and fourth diagrams in Fig. \ref{fig:diaJJ2gG2G3} starts from dimension-6 terms. 
The diagrams that contribute starting from dimension-8 condensates are presented in Fig. \ref{fig:diaJJ2gG4}. 
Since we do not consider higher dimension,  the diagrams 
in Fig.~\ref{fig:diaJJ2gG4}  could be considered in both forms: either as
local condensates or as NLC where only the leading term is taken into account.
In any case, using the leading terms of the expansion, Eq.~(\ref{eq:GGGGttttexpand}),  is useful, as it is presented in the form of the product of local scalar condensates and Lorentz tensors.

We use the NLCs and their expansions
to obtain nonperturbative contributions to OPE:  
\begin{eqnarray}\nn
	\Pi^{\pm}_\text{4}&=&\pm 4\alpha_S\va{\alpha_SG^2}\,,~~~
	\\\nn
	\Pi^{\pm}_{6}&=&\frac{8\alpha_S^2}{Q^2}\left(\pm\va{gG^3}+\frac13\va{J^2} \right)\,,\\\nn
	\Pi^{\pm}_\text{8}&=&\frac{2\alpha_S}{\pi Q^4}
	\left(2 G^8_{34}-G^8_{12}\pm 12G^8_{34}\pm G^8_{56} \right)\,,
\end{eqnarray}
where the notation for dimension-8 condensates $G^8_i$ can be found in Eq.~(\ref{eq:G8}) and (\ref{eq:G8add}), and $Q^2=-q^2$.
Our result agrees with~\cite{Novikov:1979ux,Novikov:1979va} and provides
additional contributions: dimension-6 four-quark condensate  $\va{J^2}$ and dimension-8 mixed quark-gluon condensates $G^8_{56}$. They were omitted in earlier studies~\cite{Novikov:1979ux,Novikov:1979va} as they were expected to give a minor contribution.
The details and procedures of calculation can be found in Appendices \ref{sec:subNLCaplication} and \ref{sec:subDia2}. 
There, the contributions from each diagram are given for dimension-6 and dimension-8 orders.
The case of applying the NLC expansion is discussed in detail in Appendix~\ref{sec:subDia2} using the example of diagram (d) in  Fig.~\ref{fig:diaJJ2gG2G3}.

\subsection{Three-gluon $0^{\pm +}$ glueballs }
Here we revisit OPE of the correlator used for QCD SR for the three-gluon $0^{\pm +}$ glueballs~\cite{Latorre:1987wt,Hao:2005hu}:
\begin{eqnarray}\label{eq:correlator} 
	\tilde\Pi^{\pm}(q) &=& i\int\!\! d^4x\, e^{iqx}
	\va{T \{J_3^{\pm}(x)J_3^{\dagger\pm}(0)\}}
	\\\nn
	&=&\tilde\Pi_{\text{LO}}(Q^2)+\sum_{k=1}^{5}\tilde\Pi_{k}^\pm(Q^2)+\cdots\,,
\end{eqnarray}
where the subscript $k$ numerates the NLC-based terms $\tilde\Pi_k^\pm$
and the currents are defined as follows:
\begin{eqnarray}\nn
	J_3^{+}(x)&=&g_s^3 f^{abc} 
	G^{a}_{\mu\nu}(x)
	G^{b}_{\nu\ro}(x)
	G^{c}_{\ro\mu}(x)\,,\\\nn
	J_3^{-}(x)&=&g_s^3 f^{abc} 
	\tilde G^{a}_{\mu\nu}(x)
	\tilde G^{b}_{\nu\ro}(x)
	\tilde G^{c}_{\ro\mu}(x)\,,
\end{eqnarray}
where the dual tensor 
$\tilde G^a_{\mu\nu}=i\epsilon_{\mu\nu\al\be}G_{\al\be}^a/2$.
We suggest another identical expression for a negative parity current that is easier in use:
\begin{eqnarray}\nn
	J_3^{-}(x)=g_s^3 f^{abc} 
	G^{a}_{\mu\nu}(x)
	G^{b}_{\nu\ro}(x)
	\tilde G^{c}_{\ro\mu}(x)\,.
\end{eqnarray}
Note that any other similar construction of gluon field strength tensors  
and dual tensors will be identical to $J_3^\pm$, see discussion in~\cite{Pimikov:2017bkk}, where the currents were constructed  
using helicity formalism.
We recalculated the leading perturbative term, the dimension-4 and the dimension-6 terms of OPE for two current correlators.
The leading perturbative term $\tilde\Pi_{\text{LO}}(Q^2)$ is depicted by 
the first diagram in Fig.~\ref{fig:diaJJtopG4}, the dimension-4 term comes only from the second diagram in Fig.~\ref{fig:diaJJtopG4}.
Apart from the first diagram, all diagrams presented in Fig.~\ref{fig:diaJJtopG4} contribute to the dimension-6 terms and terms of higher dimensions.
Using expansions of NLCs, we got the following result for contributions to OPE in dimension-6 order:
\begin{eqnarray}\nn
	\tilde\Pi_{\text{LO}}(Q^2) &=& -\frac{N_c^2C_F}{5\cdot 8}\alpha_s^3Q^8 \ln \frac{Q^2}{\mu^2}\,,\\\nn
	\tilde\Pi_{1}^\pm(Q^2) &=&   \pm 6\pi N_c \alpha_s^2 Q^4\va{\alpha_sG^2}
	\\\nn &&
	-6\pi N_f C_F\alpha_s^3\va{\bar qq}^2Q^2 \ln \frac{Q^2}{\mu^2} \,,\\\nn
	\tilde\Pi_{2}^\pm(Q^2) &=&  \mp (9/4)   N_c \alpha_s^2  \va{g^3G^3} Q^2   \,,\\\nn
	\tilde\Pi_{3}^\pm(Q^2) &=&  \pm 5 (9/4) N_c \alpha_s^2  \va{g^3G^3} Q^2  \,,\\\nn
	\tilde\Pi_{4}^\pm(Q^2) &=&  \mp      9  N_c \alpha_s^2  \va{g^3G^3} Q^2  \ln \frac{Q^2}{\mu^2}\,,\\\nn
	\tilde\Pi_{5}^\pm(Q^2) &=&  -\tilde\Pi_{4}^\pm(Q^2)\,,
\end{eqnarray}
where $\mu$ is the renormalization scale, $N_c$ is the number of colors, $N_f$ is the number of flavors and $C_F$ is the Casimir operator in the fundamental representation.
Our result has the same properties as the results  in~\cite{Novikov:1979ux,Novikov:1979va,Pimikov:2016pag}.
Namely, the dimension-4 and dimension-6 gluon condensate
terms have the same absolute value for both parities but different signs. 
The dimension-8 terms are expected to have the same property but only for the case of (anti-)self-dual gluon fields that can also be observed in the results for other glueball states ~\cite{Novikov:1979ux,Novikov:1979va,Pimikov:2017bkk}.

For the  three-gluon $0^{++}$ glueball, our expression for the dimension-4 term $\va{\alpha_sG^2}$ agrees with ~\cite{Latorre:1987wt}, 
the dimension-6 term, however, differs from~\cite{Latorre:1987wt}.
In the case of the  $0^{-+}$ glueball, our result for the dimension-4 contribution is the same in absolute value but has the opposite sign compared to~\cite{Hao:2005hu}.
The dimension-6 terms have 
partial correspondence and partial agreement with the terms given in~\cite{Hao:2005hu}.
Note that we take into account not only the first term of expansion but all necessary terms, Eq.~(\ref{eq:FPGexp}), for the gluon field and its strength tensor; this causes a difference for the $\tilde{\Pi}_1^-$ contribution. 
The difference for the $\tilde\Pi_3^-$ term, 
denoted by diagram (d) in Fig.~\ref{fig:diaJJtopG4},
arises from the gauge-fixing term that makes the background field vertex
different from the quantum three-gluon vertex, see Figure~\ref{fig:backgroundRules}.
The diagram (f) term $\tilde\Pi_5^-$ seems to be omitted in~\cite{Hao:2005hu}.
The $\tilde{\Pi}_2^-$ and $\tilde{\Pi}_4^-$ contributions are the same as 
in~\cite{Hao:2005hu}.

\section{Conclusions}
\label{sec:conclusions}
In this work, we have developed a new scheme to calculate OPE of vacuum correlators. 
We suggest using precalculated expansions
for NLC as intermediate to simplify calculations, where the final results for OPE are given 
by truncated series that include only local condensates up to dimension-8.
This way of using NLCs is different from the standard applications~\cite{Gromes:1982su,Mikhailov:1988nz,Bakulev:2001pa},
where condensates are considered in nonlocal form and correlator's OPE is partially resummed.
Using the procedures elaborated in~\cite{Novikov:1983gd,Grozin:1994hd},
we have obtained the full set of NLC expansions needed for calculations
of gluon condensate contributions to correlators OPE up to dimension-8 order.

The background field approach is revisited for OPE of glueball correlators.
We discuss some calculation issues which are crucial for hadron parameter evaluation within QCD SR:
(i) one of the issues is the misusing of perturbative gluon vertices that differ from the vertices for interaction of the vacuum gluon with the quantum gluon field due to the gauge fixing term;
(ii) another problem could be caused by using the incomplete Taylor expansion, Eq.~(\ref{eq:FPGexp}) -- 
all vacuum gluon fields 
and vacuum gluon field strength tensors should be expanded up to the order that could contribute to the final expression of OPE in goal dimension order;
(iii) the third issue is related to the gluon propagator expansion in the background field. 
The precalculated expressions for propagator in momentum space, Eq.~(\ref{eq:D4}), could not give  a full answer when the calculation of the coefficient function requires a derivative of the propagator in the configuration space, as in the case of glueball current correlators~\cite{Novikov:1983gd}.
In the light of discussed issues and obtained results, the glueball QCD SRs need reconsideration.

The usage of NLC expansions avoids the above problems and can be a good alternative to using precalculated expansion of the propagator in the background field.
The derived NLC expansions are universal and can be used in various applications.
Although,  only gluon condensates and their applications are considered here, the same scheme is applicable for quark and quark-gluon NLCs.
We provide the scheme for applying the NLC expansions and demonstrate it on four  correlators used in glueball state studies.
Using NLC expansions, we confirm OPEs~\cite{Novikov:1979ux,Novikov:1979va} for two-gluon $0^{\pm +}$ glueball currents and calculate
additional contributions coming from the dimension-6 four-quark condensate $\va{J^2}$ and dimension-8 mixed quark-gluon condensates 
$\va{\Tr J_\mu G_{\mu\nu}J_\nu}$ and $\va{\Tr J_\la [D_\la G_{\mu\nu},G_{\mu\nu}]}$.  
The OPE used for the three-gluon $0^{\pm +}$ glueballs~\cite{Latorre:1987wt,Hao:2005hu} are revisited up to dimension-6 order, and the corrected expressions are provided.

We would like to thank  S. Mikhailov, S. Narison, and A. Zhevlakov for stimulating discussions and useful remarks.

\begin{appendix}
	\appendix
	\section{Feynman rules}
	\label{sec:subNLCaplication}
	
\begin{figure*}[t!]	
	\centerline{
		\includegraphics[width=0.6\textwidth]{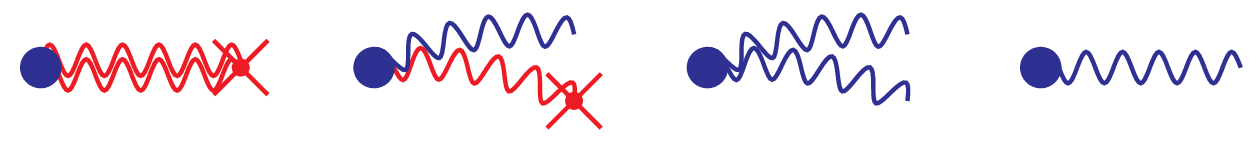}	
	}
	\caption{ 
		\label{fig:dia3L2q}
		The diagrammatic representation for the terms of the gluon field strength tensor in the glueball current. The double red line is used for the vacuum gluon field strength tensor $G^a_{\mu\nu}$. The single red line with the cross at the end denotes the vacuum gluon field. The single blue line denotes the quantum gluon field. The blue lines are part of the coefficient function and the red lines form a condensate.}
\end{figure*}

	In this appendix, the mathematical notation for the diagrams depicted in Section \ref{sec:usage} are given. 
	The right and left shaded blobs in the diagrams shown in Fig. \ref{fig:diaJJ2gG2G3}, Fig. \ref{fig:diaJJ2gG4}, and Fig.~\ref{fig:diaJJtopG4}
	represent the glueball currents with the Feynman rules provided in Fig.~\ref{fig:glueVertRules}. 
	The gluon field strength tensor $\bar G^{a}_{\mu\nu}$ in the glueball currents has
	contributions from two components of the gluon field
	$\bar A^a_\mu=A^a_\mu+a^a_\mu$,
	the perturbative quantum gluon field $a^a_\mu$, and the background field $A^a_\mu$:
\begin{eqnarray}\nn
		\bar G^{a}_{\mu\nu}(x)&=&G^{a}_{\mu\nu}(x)
		+gf^{abc}\left(A^b_\mu(x) a^c_\nu(x)-A^b_\nu(x) a^c_\mu(x)\right)
		\\\nn
		&& +gf^{abc}a^b_\mu(x) a^c_\nu(x)
		+\left(\partial_\mu a^a_\nu(x)-\partial_\nu a^a_\mu(x)\right)\,,
\end{eqnarray}
	where $G^{a}_{\mu\nu}(x)$ is the vacuum gluon field strength tensor.
	The graphical notation for these contributions is shown in Fig. \ref{fig:dia3L2q}.
	There are two types of lines: the single line represents gluon fields; the double line depicts the gluon field strength tensor. The red color and the cross at the end of the line denote the soft part that forms a vacuum condensate. 
	The blue color of the line denotes the hard part of the diagram.   
	The nonabelian part of the gluon field strength tensor contributes to diagram (b) in Fig.~\ref{fig:diaJJ2gG2G3} and diagram (e) in Fig. \ref{fig:diaJJ2gG4}, where one of the gluon fields goes to the vacuum and the other field becomes part of the gluon propagator. 
\mycomment{
	\begin{table*}[t] 	\caption{\label{fig:glueVertRules}
			The Feynman rules for glueball current vertices in the momentum representation. The single line represents the gluon field, while the double line depicts the gluon field strength tensor. The cross at the end of the single or double line tells that the given gluon field goes to vacuum and the field is part of a condensate.
		}
		\begin{center}{
				\begin{tabular}{cc}\hline\hline
					Glueball vertex & Feynman rule notation\\					\hline
					\begin{minipage}{0.3\textwidth}
						\vskip 1mm
						\begin{overpic}[width=0.4\textwidth
							]{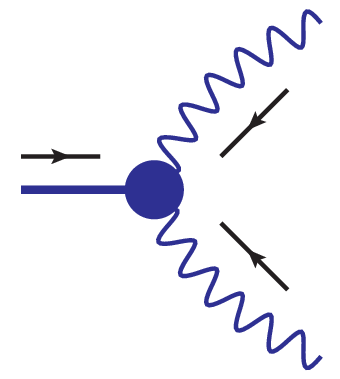}
							\put (65,60) {$p_1,~\alpha,~a_1$}
							\put (65,5) {$p_2,~\beta,~a_2$}
							\put ( 5,50) {$q$}			
						\end{overpic}
						\vskip 1mm
					\end{minipage}
					& 
					~~$\alpha_S\delta^{a_1a_2}T^P_{\mu\nu}\mathbb{A}(\mu_1,\nu_1)
					\mathbb{A}(\mu_2,\nu_2)
					ip_{1\mu_1} ip_{2\mu_2}g_{\nu_1\alpha}g_{\nu_2\beta}
					\delta(p_1+p_2+q)$~~
					\\
					\begin{minipage}{0.3\textwidth}		 	 	
						\vskip 1mm
						\begin{overpic}[width=0.4\textwidth
							]{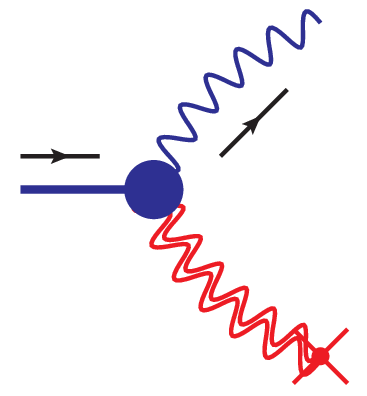}
							\put (65,60) {$q,~\alpha,~a_1$}
							\put (65,5) {$\mu_2\nu_2,~a_2$}
							\put ( 5,50) {$q$}			
						\end{overpic} 
					    \vskip 1mm
					\end{minipage}
					& 
					$\alpha_S\delta^{a_1a_2}T^P_{\mu\nu}\mathbb{A}(\mu_1,\nu_1)
					iq_{\mu_1} g_{\nu_1\alpha} $
					\\
					\begin{minipage}{0.3\textwidth}
						\vskip 1mm
						\includegraphics[width=0.4\textwidth]{glueball-vertex-GaA.eps}
						\begin{overpic}[width=0.4\textwidth
							]{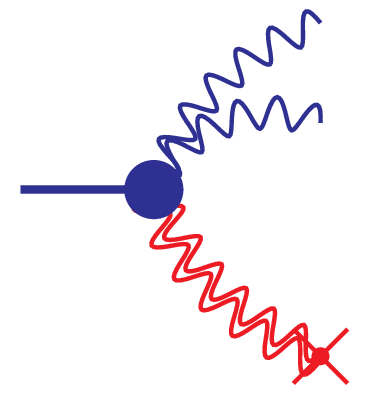}
							\put (65,60) {$\mu_1,b$}
							\put (65,45) {$\nu_1,c$}
							\put (65,5) {$\mu_2\nu_2,a_2$}
						\end{overpic} 
					\vskip 1mm
					\end{minipage}
					\hspace{5mm}
					& 
					$\alpha_Sgf^{a_2bc}T^P_{\mu\nu} $
					\\\hline\hline
				\end{tabular}
			}
		\end{center}
	\end{table*}
\begin{table*}[t] 	\caption{\label{fig:backgroundRules}
		The Feynman rules for the background field vertices.  
		We use the following matrix notation for the vacuum gluon field $A_\mu^{ab}=g f^{acb}A^c_\mu(y)$ and its strength tensor $G^{ab}_{\mu\nu}=g f^{acb}G^c_{\mu\nu}(y)$, where $y$ is the interaction point.
	}
	\begin{center}{
			\begin{tabular}{cc}\hline\hline
				Vertex & Feynman rule notation\\					\hline
				\begin{minipage}{0.3\textwidth}
					\vskip 1mm
					~~~~~\begin{overpic}[width=0.7\textwidth
						]{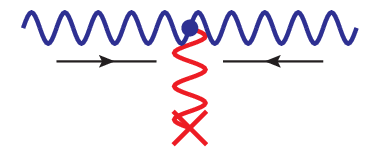}
						\put (-10,25) {$\alpha,a$}
						\put (105,25) {$\beta,b$}
						\put (70,14) {$k_2$}
						\put (25,14) {$k_1$}
					\end{overpic}
				\end{minipage}
				& 
				\begin{minipage}{0.7\textwidth}
					$V^{ab}_{1\al\be}=A^{ab}_\mu(k_2-k_1)_\mu g_{\alpha\beta}+A^{ab}_\alpha(k_1+k_2)_\beta-A^{ab}_\beta(k_1+k_2)_\alpha +iG^{ab}_{\alpha\beta}$
				\end{minipage}	
				\\
				\begin{minipage}{0.3\textwidth}
					~~~~~\begin{overpic}[width=0.7\textwidth 
						]{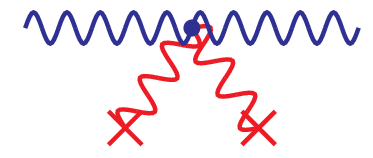}
						\put (-10,25) {$\alpha,a$}
						\put (105,25) {$\beta,b$}
					\end{overpic}
				\end{minipage}
				& 
				\begin{minipage}{0.7\textwidth}
					$V^{ab}_{2\al\be}=i(A^{ac}_\mu A^{cb}_\mu g_{\alpha\beta}+[A_\alpha, A_\beta]^{ab})$
				\end{minipage}
				\\\hline\hline
			\end{tabular}
		}
	\end{center}
\end{table*}
}

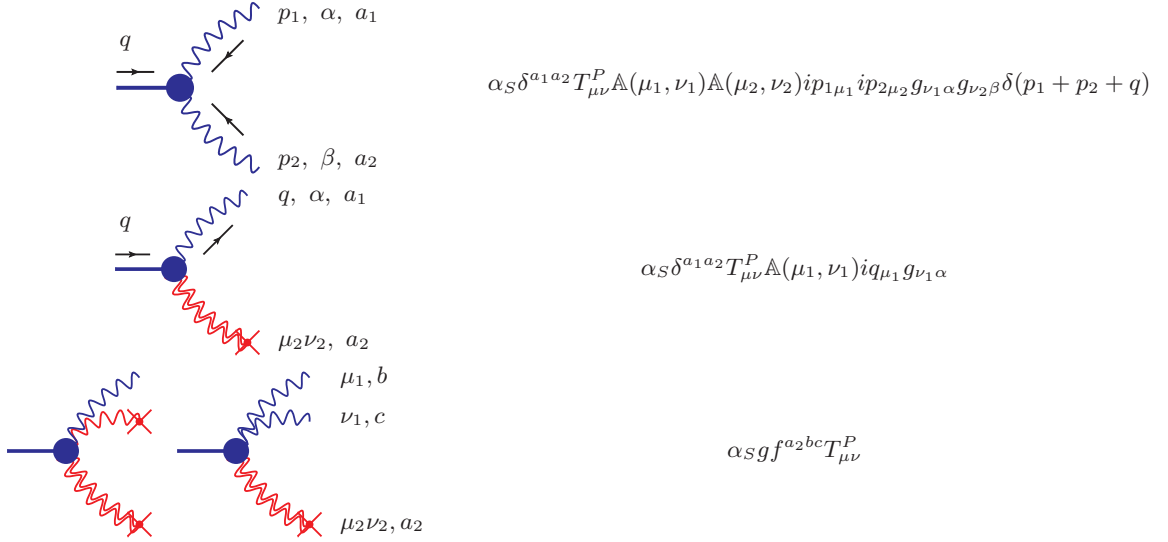
\begin{figure*}[t!]	
	\begin{center}{
			\begin{tabular}{cc}
				\begin{minipage}{0.3\textwidth}
					\vskip 1mm
					\begin{overpic}[width=0.4\textwidth
						]{glueball-vertex-G0G0.eps}
						\put (65,60) {$p_1,~\alpha,~a_1$}
						\put (65,5) {$p_2,~\beta,~a_2$}
						\put ( 5,50) {$q$}			
					\end{overpic}
					\vskip 1mm
				\end{minipage}
				& 
				~~~~~~~~$\alpha_S\delta^{a_1a_2}T^P_{\mu\nu}\mathbb{A}(\mu_1,\nu_1)
				\mathbb{A}(\mu_2,\nu_2)
				ip_{1\mu_1} ip_{2\mu_2}g_{\nu_1\alpha}g_{\nu_2\beta}
				\delta(p_1+p_2+q)$~~
				\\
				\begin{minipage}{0.3\textwidth}		 	 	
					\vskip 1mm
					\begin{overpic}[width=0.4\textwidth
						]{glueball-vertex-GG0.eps}
						\put (65,60) {$q,~\alpha,~a_1$}
						\put (65,5) {$\mu_2\nu_2,~a_2$}
						\put ( 5,50) {$q$}			
					\end{overpic} 
					\vskip 1mm
				\end{minipage}
				& 
				$\alpha_S\delta^{a_1a_2}T^P_{\mu\nu}\mathbb{A}(\mu_1,\nu_1)
				iq_{\mu_1} g_{\nu_1\alpha} $
				\\
				\begin{minipage}{0.3\textwidth}
					\vskip 1mm
					\includegraphics[width=0.4\textwidth]{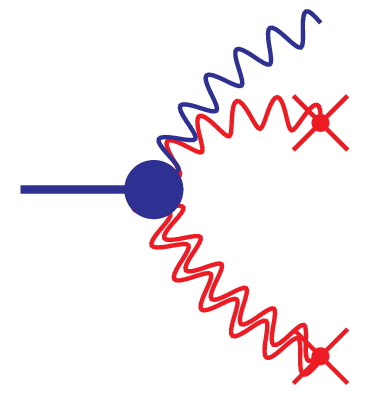}
					\begin{overpic}[width=0.4\textwidth
						]{glueball-vertex-Gaa.eps}
						\put (65,60) {$\mu_1,b$}
						\put (65,45) {$\nu_1,c$}
						\put (65,5) {$\mu_2\nu_2,a_2$}
					\end{overpic} 
					\vskip 1mm
				\end{minipage}
				\hspace{5mm}
				& 
				$\alpha_Sgf^{a_2bc}T^P_{\mu\nu} $
			\end{tabular}
		}
	\end{center}
	\caption{ \label{fig:glueVertRules}
		The Feynman rules for contributions to glueball current vertices in the momentum representation. The single line represents the gluon field, while the double line depicts the gluon field strength tensor. The cross at the end of the single or double line tells that the given gluon field goes to vacuum and the field is part of a condensate.
	}
\end{figure*}
	
	The graphic notation and definitions for the vertices of the quantum field and background field interactions
	are provided in Fig.~\ref{fig:backgroundRules}. 
	The vertices are given in terms of the operators of the vacuum gluon field and its strength tensor at the point of interaction. 
	These operators become part of NLCs. 
	Otherwise, these operators need to be expanded to the Taylor series,  Eq.~(\ref{eq:FPGexp}).
	
	Calculation in the  configuration space requires integration over the coordinates of interaction points, see, for example, diagrams (e), (f), or (g) in Fig.~\ref{fig:diaJJ2gG4}.
	We consider differentiating easier than integrating; therefore, 
	we work in the momentum representation with the following definition for 
	the coordinates that appear in the obtained expansion of nonlocal condensates.
	For the coordinate $x$, where one of the currents in the correlator
	is located, we have
	\begin{equation}\label{eq:xxx}
		x_{\rho_1}x_{\rho_2}\ldots x_{\rho_n} \to
		\frac{\partial}{i\partial q_{\rho_1}}
		\frac{\partial}{i\partial q_{\rho_2}}\ldots 
		\frac{\partial}{i\partial q_{\rho_n}}\,.
	\end{equation}
	For any arbitrary coordinate $y$ in NLC expansion differing from $x$ we apply
	\begin{equation}\label{eq:yyy}
		y_{\rho_1}y_{\rho_2} \ldots y_{\rho_k}\to \int d^4l e^{-ily} 
		\left(
		\frac{\partial}{i\partial l_{\rho_1}}
		\frac{\partial}{i\partial l_{\rho_2}}\ldots 
		\frac{\partial}{i\partial l_{\rho_k}}
		\delta^4(l)
		\right)\,,
	\end{equation}
	where additional coordinate $y$ is a point of interaction with the background field and $l$ is auxiliary momentum running from point $0$ to $y$.
	
	The general procedure for OPE calculation in our approach is the following.
	Using Wick's theorem, we obtain all possible contributions and depict the corresponding Feynman diagrams. 
	We do not perform Taylor and tensor expansions of the vacuum fields,
	but use NLC expansions to get a truncated OPE.
	As the  next step, we apply the following Feynman rules and expressions to:
	the background field vertices, Figure \ref{fig:backgroundRules};
	current vertices for the two-gluon glueball, figure \ref{fig:glueVertRules};
	two-gluon NLC expansion, Eq. (\ref{eq:GGnlc});
	three-gluon NLC expansion, Eq. (\ref{eq:GGGnlc});
	four-gluon NLC expansion, Eq. (\ref{eq:GGGGttttexpand}) and Eq. (\ref{eq:GGGGffExpand});
	coordinates from NLC expansions, Eq.~(\ref{eq:xxx}) and Eq.~(\ref{eq:yyy});
	and free gluon propagator (without background field corrections). 

\begin{figure*}[t!]
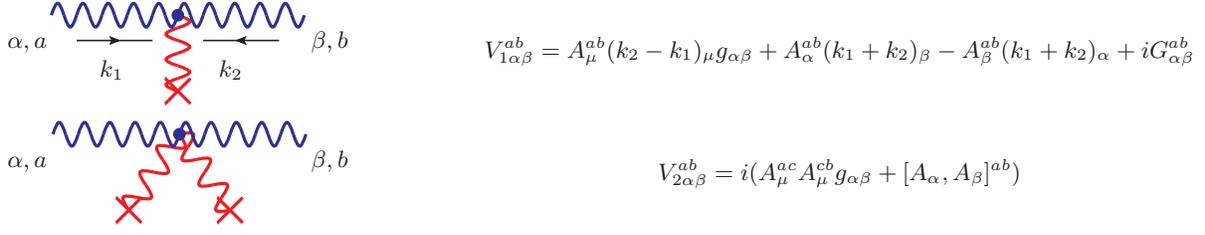
	
	\begin{center}{
			\begin{tabular}{cc}
				\begin{minipage}{0.3\textwidth}
					\vskip 1mm
					~~~~~\begin{overpic}[width=0.7\textwidth
						]{background-feynrules-vertex-aAa-newStyle.eps}
						\put (-10,25) {$\alpha,a$}
						\put (105,25) {$\beta,b$}
						\put (70,14) {$k_2$}
						\put (25,14) {$k_1$}
					\end{overpic}
				\end{minipage}
				& 
				\begin{minipage}{0.7\textwidth}
					$V^{ab}_{1\al\be}=A^{ab}_\mu(k_2-k_1)_\mu g_{\alpha\beta}+A^{ab}_\alpha(k_1+k_2)_\beta-A^{ab}_\beta(k_1+k_2)_\alpha +iG^{ab}_{\alpha\beta}$
				\end{minipage}	
				\\
				\begin{minipage}{0.3\textwidth}
					~~~~~\begin{overpic}[width=0.7\textwidth 
						]{background-feynrules-vertex-aAAa-newStyle.eps}
						\put (-10,25) {$\alpha,a$}
						\put (105,25) {$\beta,b$}
					\end{overpic}
				\end{minipage}
				& 
				\begin{minipage}{0.7\textwidth}
					$V^{ab}_{2\al\be}=i(A^{ac}_\mu A^{cb}_\mu g_{\alpha\beta}+[A_\alpha, A_\beta]^{ab})$
				\end{minipage}
			\end{tabular}
		}
	\end{center}
	\caption{ 
		\label{fig:backgroundRules}
		The Feynman rules for the background field vertices.  
		We use the following matrix notation for the vacuum gluon field $A_\mu^{ab}=g f^{acb}A^c_\mu(y)$ and its strength tensor $G^{ab}_{\mu\nu}=g f^{acb}G^c_{\mu\nu}(y)$, where $y$ is the interaction point.}
\end{figure*}

	\section{Example of NLC-based calculation}
	\label{sec:subDia2}
	
	In this section, we provide the technical details of the OPE calculation of the glueball current correlator $\Pi^{P}(q)$, Eq.~(\ref{eq:correlatorP}), based on grouping the contributions in NLCs
	and using precalculated expressions for NLC expansions. 
	As an example, the diagram (d) term $\Pi^\pm_{\text{NLC-}2}$ depicted in Fig. \ref{fig:diaJJ2gG2G3} is considered. 
	Applying the Feynman rules given in Appendix~\ref{sec:subNLCaplication}, we have:
	\begin{eqnarray}\nn
		\Pi^\pm_{\text{NLC-}2}&=& i 4 \va{G^{a_2}_{\mu_2\nu_2}(0) V^{c_1c_2}_{1\rho_1\rho_2}(y)G^{b_2}_{\al_2\be_2}(x)}
	\\\nn&&
		\cdot\frac{g_{\bar\nu_1\rho_2}\delta^{a_1c_2}}{i (q-l)^2}
		\cdot\frac{g_{\bar\be_1\rho_1}\delta^{c_1b_1}}{i q^2}
	\\\nn&&
		\cdot \al_S \delta^{a_1a_2} T^\pm_{\mu\nu} \mathbb{A}(\mu_1,\nu_1) i(q-l)_{\mu_1}g_{\nu_1\bar\nu_1}
	\\\nn&&
		\cdot \al_S \delta^{b_1b_2} T^\pm_{\al\be} \mathbb{A}(\al_1,\be_1) i(-q_{\al_1})g_{\be_1\bar\be_1}\,,
	\end{eqnarray}
	where expression is given in the form of the product of NLC, two propagators and 
	two glueball vertices. 
	The tensors 
	$T^{+}_{\mu\nu}\equiv g_{\mu_1\mu_2}g_{\nu_1\nu_2}$, and
	$T^{-}_{\mu\nu}\equiv i\epsilon_{\mu_1\nu_1\mu_2\nu_2}/2$ define the glueball current.
	
	The expression for the correlator is given in the momentum space where the coordinates $x$ and $y$ are considered as differential operators defined 
	in Eqs. (\ref{eq:xxx}) and (\ref{eq:yyy}). Next, we consider the example of the application of these operators.
	The condensate part is given by
	\begin{eqnarray}\nn
		\va{G^a_{\mu_2\nu_2}(0)A^{b}_{\rho_3}(y)G^c_{\alpha_2\beta_2}(x)}&&
	\\\nn	
	&&\hspace{-35mm}=
		y_{\rho_0}\!\int_0^1\!\!dt~t\va{G^a_{\mu_2\nu_2}(0)G^{b}_{\rho_0\rho_3}(ty)G^c_{\alpha_2\beta_2}(x)}
		\\\label{eq:PiDia2}
	&&\hspace{-35mm}=
		\frac{y_{\rho_0}}{2}\va{G^a_{\mu_2\nu_2}(0)G^{b}_{\rho_0\rho_3}(0)G^c_{\alpha_2\beta_2}(0)}
	\\\nn && \hspace{-35mm} ~~~
		+\ldots ~,
	\end{eqnarray}
	where, in the third line, we keep only the leading term in the expansion of the gluon field to demonstrate
	the calculation of the dimension-6 term.
	Then, for leading order, we replace the condensate by its expansion, Eq.~(\ref{eq:GGGnlc}), and replace the coordinates by their representations in the momentum space, Eqs. (\ref{eq:yyy}),
	which take a shorter form here
	\begin{equation}\nn
		y_\rho\to\int d^4l \,e^{-ily} \frac{\partial}{i\partial l_\rho}\delta^4(l)\,.
	\end{equation}
	To consider the next term that has dimension-8, we apply the three-gluon NLC expansion given in Eq.~(\ref{eq:GGGnlc}).  
	One could choose not to use the NLC expansion, then one should take the first three terms of Taylor expansion, see Eq. (\ref{eq:FPGexp}), for all gluon field strength tensors in Eq. (\ref{eq:PiDia2}).

	Applying Eq. (\ref{eq:FS-integral}) and the expansion of the three-gluon NLC given  in Eq. (\ref{eq:GGGnlc}),  one gets
\begin{widetext}
	\begin{eqnarray}\nn
		\Pi^\pm_{\text{NLC-}2}&=&\frac{-\alpha_S}{6\pi}
		T^\pm_{\mu\nu}T^\pm_{\al\be} 
		\mathbb{A}({\mu_1,\nu_1}) \mathbb{A}({\mu_2,\nu_2}) 
		\mathbb{A}({\al_1,\be_1}) \mathbb{A}({\al_2,\be_2})
		\mathbb{A}({\rho_0,\rho_3})
		\int\limits_0^1\!\! dt
		\left(\frac 12 \mathbb{A}
		\sum\limits_{i=0}^{7}
		\Gamma_{i}^{(abc)}M_{i}(x_a,x_b,x_c)
		\right)
		\\\label{eq:dia2aux}
		&&\cdot 
		\left(
		-ity_{\rho_0} V_{1\be_1\nu_1\rho_3}(-q,q-l,l)
		-g_{\rho_0\be_1}g_{\rho_3\nu_1}\delta(t-1)
		\right)
		\cdot 
		\frac{q_{\al_1}(q-l)_{\mu_1}}{q^2(q-l)^2}\,,
	\end{eqnarray}
\end{widetext}
	where part of the background field vertex is given by
	\begin{eqnarray}\nn
		V_{1\al_1\al_2\al_3}(k_1,k_2,k_3)&&
	\\\nn   && \hspace{-20mm}
		=
		g_{\al_1\al_2}(k_1-k_2)_{\al_3}
		-g_{\al_1\al_3}(k_1-k_2)_{\al_2}
	\\\nn  &&  \hspace{-20mm} ~~
		+g_{\al_2\al_3}(k_1-k_2)_{\al_1}\,,
	\end{eqnarray}
	and the coordinates of the scalar functions and tensors are $x_1=0$, $x_2=yt$, $x_3=x$.
	The master tensors $\Gamma_{i}$ carry the Lorentz indices $\mu_2\nu_2\al_2\be_2\rho_0\rho_3$, see Eq. (\ref{eq:tensorsrArB}). 
	From Eq. (\ref{eq:dia2aux}) one gets 
	\begin{eqnarray}\nn
		\Pi^\pm_\text{NLC-2} &=& \pm\frac{12\alpha_S^2\va{gG^3}}{Q^2} \\\nn
		&& +\frac{2\alpha_S}{3\pi Q^4}
		\left(\mp 12G^8_{34}\pm 3 G^8_{56} +2 G^8_{34}-G^8_{12} \right)\,.
	\end{eqnarray}

	OPE of the scalar and pseudoscalar glueballs current correlator $\Pi^{\pm}$, defined in Eq.~(\ref{eq:correlatorP}),
	is given by the sum of NLC-based terms $\Pi^\pm_{\text{NLC-}k}$ depicted in Fig. \ref{fig:diaJJ2gG2G3} and Fig. \ref{fig:diaJJ2gG4}.
	We denote the dimension-$n$ contribution of the given NLC-based diagram by $\Pi^\pm_{nk}$
	and consider the contributions up to dimension-8 order:
	\begin{eqnarray}\nn
		\Pi^\pm_{\text{NLC-}k}=\sum\limits_{n=d_k}^8 \Pi^\pm_{nk}\,,
	\end{eqnarray}
	where $d_k$ is the dimension of the $k$-th NLC-based term.
	Only diagram (b) contributes to leading dimension-4: 
	\begin{eqnarray}\nn
		\Pi^{\pm}_\text{4}=\Pi^{\pm}_\text{40}=
		\pm 4\alpha_S\va{\alpha_SG^2}\,.
	\end{eqnarray}
	There are three diagrams that contribute to dimension-6 order: 
	\begin{eqnarray}\nn
		\Pi^{\pm}_{6}&=&\sum\limits_{k=0}^{2}\Pi^{\pm}_{6k}
		=\frac{8\alpha_S^2}{Q^2}\left(\pm\va{gG^3}+\frac13\va{J^2} \right)\,,\\\nn
		\Pi^{\pm}_{60}&=&\frac{8\alpha_S^2}{3Q^2}\va{J^2}\,,~~~
		\Pi^{\pm}_{61}=\mp\frac{4\alpha_S^2\va{gG^3}}{Q^2}\,,
		\\\nn
		\Pi^{\pm}_{62}&=&\pm\frac{12\alpha_S^2\va{gG^3}}{Q^2}\,.
	\end{eqnarray}
	All considered diagrams can contribute to dimension-8 order:
	\begin{eqnarray}\label{eq:piG4sum}
		\Pi^{\pm}_{8}&=&\frac{2\alpha_S}{\pi Q^4}\sum\limits_{k=0}^{5}\Pi^{\pm}_{8k}
	\\\nn
		&=&\frac{2\alpha_S}{\pi Q^4}
		\left(2 G^8_{34}-G^8_{12}\pm 12G^8_{34}\pm G^8_{56} \right)\,,
	\end{eqnarray}
	where the $k$-th diagram contribution is denoted by $\Pi^{\pm}_{8k}$ and 
	given in Table \ref{tab:condG4terms0pmplusvert} by local condensates. 
	Diagrams (b), (c), and (e) give a zero contribution for  the space dimension $d=4$.

	\begin{table}[h]
		\caption{\label{tab:condG4terms0pmplusvert}
			The dimension-8 contribution to each diagram $\Pi_{8k}^{\pm}$, where $k$ is number of the diagram, as depicted in Fig. (\ref{fig:diaJJ2gG2G3}) and (\ref{fig:diaJJ2gG4}), see Eq.~(\ref{eq:piG4sum}).
		}
		\begin{center}
			\renewcommand{\arraystretch}{1.3}
			\begin{tabular}{ccc}\hline\hline
				$k$ &  $\Pi^+_{8k}$ &  $\Pi^-_{8k}$	\\\hline 
				0 & $O(d-4)$ & $O(d-4)$ \\
				1 & $O(d-4)$ & $O(d-4)$ \\
				2 & $-\frac{1}{3}(G^8_{12}+10G^8_{34}-3G^8_{56})$~ &  
				    ~$-\frac{1}{3}(G^8_{12}-14G^8_{34}+3G^8_{56})$ \\
				3 & $O(d-4)$ & $O(d-4)$ \\
				4 & $18G^8_{34}$ &  $-G^8_{12}-16G^8_{34}$ \\
				5 & $-2(G^8_{12}+G^8_{34})/3$ &  $(G^8_{12}+4G^8_{34})/3$ \\
				sum & $  14G^8_{34}-G^8_{12}+G^8_{56}$ &  
				$-(10G^8_{34}+G^8_{12}+G^8_{56})$ \\\hline\hline
			\end{tabular}
	\end{center}
\end{table}

\end{appendix}
 


\end{document}